\documentclass[prl,superscriptaddress,showpacs,twocolumn]{revtex4-1}

\usepackage[normalem]{ulem}
\usepackage{graphicx}
\usepackage{color}
\usepackage{verbatim}
\usepackage{subfigure}
\usepackage{amssymb}
\usepackage{amsmath}	
\usepackage{comment}

\newcommand{\sone}[0]{\delta_{ij}\delta_{kl}}
\newcommand{\stwo}[0]{(\delta_{ik}\delta_{jl}-\epsilon_{ik}\epsilon_{jl})}
\newcommand{\sthree}[0]{\epsilon_{ij}\epsilon_{kl}}
\newcommand{\sfour}[0]{(\epsilon_{ik}\delta_{jl}+\epsilon_{jl}\delta_{ik})}
\newcommand{\sfive}[0]{(\epsilon_{ik}\delta_{j\ell}-\epsilon_{j\ell}\delta_{ik}+\epsilon_{ij}\delta_{k\ell}+\epsilon_{k\ell}\delta_{ij})}
\newcommand{\ssix}[0]{(\epsilon_{ik}\delta_{j\ell}-\epsilon_{j\ell}\delta_{ik}-\epsilon_{ij}\delta_{k\ell}-\epsilon_{k\ell}\delta_{ij})}

\newcommand{\tcor}[2]{\left\lbrace#1\right\rbrace_{#2}}

\newcommand{\brak}[1]{\left\langle#1\right\rangle}

\usepackage{amsthm}

\usepackage{bm}

\theoremstyle{plain}
\newtheorem{theorem}{Theorem}

\newcommand{\im}[0]{\text{i}}

\begin{document}

\title{Time reversal symmetry breaking in two-dimensional\\
non-equilibrium viscous fluids}

\author{Jeffrey M. Epstein}
\email{epstein@berkeley.edu}
\affiliation{Department of Physics, University of California, Berkeley, CA, USA}
 
\author{Kranthi K. Mandadapu}
\email{kranthi@berkeley.edu}
\affiliation{Department of Chemical and Biomolecular Engineering, University of California, Berkeley, CA, USA}
\affiliation{Chemical Sciences Division, Lawrence Berkeley National Laboratory, Berkeley, CA, USA}

\begin{abstract}
We study the rheological signatures of departure from equilibrium in two-dimensional viscous fluids with and without internal spin.
Under the assumption of isotropy, we provide the most general linear constitutive relations for stress and couple stress in terms of the velocity and spin fields. Invoking Onsager's regression hypothesis for fluctuations about steady states, we derive the Green-Kubo formulae relating the transport coefficients to time correlation functions of the fluctuating stress. In doing so, we verify the claim that one of the non-equilibrium transport coefficients, the odd-viscosity, requires time reversal symmetry breaking in the case of systems without internal spin. However, the Green-Kubo relations for systems with internal spin also show that there is a possibility for non-vanishing odd viscosity even when time reversal symmetry is preserved.
Furthermore, we find that breakdown of equipartition in non-equilibrium steady states results in the decoupling of the two rotational viscosities relating the vorticity and the internal spin. 
\end{abstract}

\maketitle

\noindent\textbf{\textit{Introduction.}} 
This paper presents the consequences of time reversal symmetry breaking at the microscale, and other signatures of non-equilibrium on emergent transport coefficients in two-dimensional viscous fluids. A motivation for this work is the recent emergence of the field of active matter, which studies systems that consume and dissipate energy at the particle scale. 
Active systems have been found to yield novel phase behaviors \cite{vicsek1995novel, gregoire2004onset, tailleur2008statistical, fily2012athermal, redner2013structure, solon2015pressure} and continuum descriptions with unusual transport behaviors \cite{simha2002hydrodynamic, takatori2014swim, takatori2017superfluid, hatwalne2004rheology,henkin2014tunable, guillamat2016probing, wan2008rectification,angelani2010geometrically,ghosh2013self, klymko2017statistical, epstein2018statistical, dasbiswas2018topological} including odd viscosity  \cite{banerjee2017odd, souslov2019topological}. They also provide insights into activity-mediated biological processes, including flows in the actin cortex \cite{nishikawa2017controlling,mayer2010anisotropies} and collective motion in swarming and growing bacterial colonies \cite{wensink2012meso,cates2010arrested,doostmohammadi2016defect}. 

Active systems with non-conservative forces at the microscale result in non-equilibrium steady states different in nature from those arising from spatial gradients in temperature, pressure, or chemical potential by means of boundary conditions. The latter class of problems has been of intense interest for over a century and may be addressed within a well-established non-equilibrium thermodynamics formalism that unifies a variety of transport processes, building on the seminal work of Onsager, Prigogine, deGroot, and Mazur \cite{Onsager1931a, Onsager1931b, Prigogine1967, Mazur1984}. This approach is based on the local equilibrium hypothesis, expressing entropy production as a bilinear form of  generalized thermodynamic forces $X_\alpha$ and fluxes $J_\alpha$, with $\alpha$ enumerating the concerned transport process. 
The fluxes and forces are then related by linear laws, 
\begin{equation}
J_\alpha=\sum_{\beta}L_{\alpha\beta}X_\beta .
\end{equation} 
The proportionality constants $L_{\alpha \beta}$, also referred to as transport coefficients, obey the celebrated Onsager reciprocal relations, $L_{\alpha \beta} = L_{\beta \alpha}$, derived by Onsager via invocation of the principle of microscopic time-reversibility (or time reversal symmetry),
and a regression hypothesis connecting the macroscopic boundary-driven gradient phenomena to fluctuations in equilibrium systems \cite{Onsager1931a,Onsager1931b}. 
These same assumptions were used by Kubo, Yokota, and Nakajima to derive another set of prominent relations, the Green-Kubo relations, relating the constants $L_{\alpha \beta}$ to integrals of the time correlation functions of the fluxes $J_\alpha$ in equilibrium systems \cite{Kubo1957, Kubo1957b}.

Active matter systems, which break time reversal symmetry at the microscale, still lack a unifying thermodynamic description for explaining emergent transport phenomena.
In this work, we study the non-equilibrium viscous transport behaviors of generic isotropic active systems in two-dimensions with and without internal spin, investigate the fluctuations in the non-equilibrium steady state, analyze the consequences of time reversal symmetry breaking on the emergent transport coefficients, and demonstrate the breakdown of Onsager's reciprocal relations by deriving Green-Kubo relations. We thus provide a first step towards developing the non-equilibrium thermodynamics formalism for transport phenomena in active media. \\

\noindent\textbf{\textit{Odd Viscosity and Time-Reversal.}} 
We begin our treatment of emergent behavior in viscous non-equilibrium fluids with a brief discussion of odd viscosity, recently proposed as a consequence of time reversal symmetry breaking in active media \cite{avron1995viscosity,Avr98,banerjee2017odd,ganeshan2017odd}. 
In general, the viscosity tensor $\eta_{ijkl}$ defines the linear relation between the stress tensor $T_{ij}$ and the velocity gradient $v_{k,l}$ in a viscous fluid, where $(\cdot)_{,i}$ indicates the spatial derivative. 
Certain symmetries of the viscosity are physically meaningful. It is well-known that the symmetry $\eta_{ijkl}=\eta_{jikl}$ enforces the symmetry of the stress tensor, while the symmetry $\eta_{ijkl}=\eta_{ijlk}$ expresses the objectivity of the stress tensor, \emph{i.e.}, its insensitivity to the antisymmetric part of the velocity gradient, reflecting the rigid body rotation of the fluid. 
Another symmetry that has recently attracted interest is $\eta_{ijkl}=\eta_{klij}$ \cite{banerjee2017odd,Avr98}. Components of the viscosity that are antisymmetric with respect to this permutation do not contribute to the stress power $T_{ij}v_{i,j}= \eta_{ijkl}v_{i,j}v_{k,l}$, and are referred to as odd viscosities. 
These have interesting hydrodynamical consequences such as transverse response to shear strain that 
have been explored in both classical \cite{ganeshan2017odd,banerjee2017odd} and quantum \cite{avron1995viscosity} fluids. In the quantum setting, the odd viscosity is expected to appear in quantum Hall fluids \cite{Avr98}.

In previous works associated with odd viscosity \cite{banerjee2017odd, Avr98}, the symmetry $\eta_{ijkl}=\eta_{klij}$ has been claimed as a necessary consequence of time reversal symmetry, on the basis of Onsager's reciprocal relations $L_{\alpha \beta} = L_{\beta \alpha}$. At first glance, it is plausible to take the symmetry $\eta_{ijkl}=\eta_{klij}$ of the viscosity to be a particular instance of Onsager reciprocity, identifying $\alpha=ij$ and $\beta=kl$, with stress being the flux of momentum and velocity gradient as the generalized thermodynamic force. 
This analogy is incorrect, as the reciprocal relations were developed for coupled thermodynamical transport processes using the entropy as a central tool \cite{Onsager1931a,Onsager1931b}. The center of mass momentum of a fluid parcel is not a thermodynamic quantity, and does not enter into any proper account of the entropy of that parcel. Thus an independent demonstration is required to prove that time reversal symmetry breaking is necessary for the observation of odd viscosity, and therefore the breakdown of the reciprocal relations for the viscosity tensor. 
This is one of the results we provide here, still using Onsager's particular insight, the regression hypothesis connecting the fluctuations in the steady state to macroscopic boundary-driven gradients in velocity.

Generalizations to other thermodynamic transport processes involving gradients in pressure, chemical potential and temperature, and the consequences of time reversal symmetry breaking on the corresponding transport coefficients and reciprocal relations will require understanding the role of entropy and entropy production in non-equilibrium steady states \cite{fodor2016far, mandal2017entropy} and will be left for future work. In this work, we focus on the viscous effects in fluids with internal structure, and study only transport processes involving gradients in dynamical quantities based on momentum and internal spin. \\

\noindent\textbf{\textit{Conservation laws.}} To study two-dimensional fluids with internal structure,
we take as fundamental dynamical fields the velocity vector $v_i$ and the scalar internal spin $m$. 
Conservation of linear and angular momentum are guaranteed by the balance equations
\begin{align}
\rho\dot{v}_i&=T_{ij,j},\label{eq:balmom}\\
\rho\dot{m}&=C_{i,i} -\epsilon_{ij}T_{ij},\label{eq:balangmom}
\end{align}
with $T_{ij}$ the stress tensor and $C_i$ the couple stress or spin flux. The dot indicates the convective or material derivative $\partial_t+v_i \partial_i$, and $\epsilon_{ij}$ is the two-dimensional Levi-Civita tensor. Such a microstructural continuum theory was proposed by Dahler and Scriven \cite{dahler1961angular,Dah63} and has been used in many contexts \cite{tsai2005chiral,van2016spatiotemporal,furthauer2012active,furthauer2013active}.

The coupling term $-\epsilon_{ij}T_{ij}$ between the linear and internal angular momentum balance equations preserves conservation of total angular momentum, with density $\mathcal{J}=\rho\mathbf{x}\times\mathbf{v}+\rho m$ ($\mathbf{x}$ is the position),
while permitting the existence of a nonvanishing antisymmetric component of the stress. This is forbidden by conservation of angular momentum in fluids without a microstructural field capable of absorbing angular momentum from the velocity field.  \\

\noindent\textbf{\textit{Constitutive relations and isotropy.}} To close the equations \eqref{eq:balmom} and \eqref{eq:balangmom} for $v_i$ and $m$, we require constitutive equations relating the stress $T_{ij}$ and couple stress $C_i$ to the fields $v_i$ and $m$. We assume these relations are linear, Galilean invariant, and contain derivatives of the fields only up to first order. The most general linear constitutive relations are then given by 
\begin{align}
T_{ij}&=\eta_{ijkl}v_{k,l}+\gamma_{ij}m+\xi_{ijk}m_{,k}\label{eq:linearmap-stress},\\
C_i&=\beta_{ijk}v_{j,k}+\kappa_im+\alpha_{ij}m_{,j} \label{eq:linearmap-couple},
\end{align}
where repeated indices are summed, and $\boldsymbol{\eta}$, $\boldsymbol{\gamma}$, $\boldsymbol{\xi}$, $\boldsymbol{\beta}$, $\boldsymbol{\kappa}$, and $\boldsymbol{\alpha}$ are linear maps.

Imposing isotropy further restricts the couplings in \eqref{eq:linearmap-stress} and \eqref{eq:linearmap-couple}. Isotropic tensors of any rank in dimension $n$ may be expressed as linear combinations of terms consisting only of the rank two Kronecker tensor $\delta_{ij}$ and the rank $n$ Levi-Civita tensor $\epsilon_{i_1\ldots i_n}$  (see Appendix I). In two dimensions, both of these are rank two, so there are no nonzero isotropic tensors of odd rank. This forbids the existence of nontrivial isotropic linear maps between tensors with ranks differing by an odd number. For instance, the couple stress $C_i$, a vector, cannot depend on the spin density $m$, a scalar, or the velocity gradient $v_{i,j}$, a rank two tensor. Similarly, the stress tensor $T_{ij}$, a rank two tensor, cannot depend on the spin gradient $m_{,i}$, a vector. Therefore, the most general isotropic constitutive equations have the form
\begin{align}
T_{ij}&=\eta_{ijkl}v_{k,l}+\gamma_{ij}m\label{eq:linearmapstress-reduc},\\
C_i&=\alpha_{ij}m_{,j}\label{eq:linearmapspin-reduc}.
\end{align}
The maps $\gamma_{ij}$ and $\alpha_{ij}$ may be expressed as
\begin{align}
\gamma_{ij}&=\gamma_1\delta_{ij}+\gamma_2\epsilon_{ij}\label{eq:gammas},\\
\alpha_{ij}&=\alpha_1\delta_{ij}+\alpha_2\epsilon_{ij}\label{eq:alphas}.
\end{align}
The viscosity tensor $\eta_{ijkl}$ is an element of the six-dimensional space of isotropic rank four tensors in two dimensions (see Appendix I). An orthogonal basis $\mathbf{s}^{(\alpha)}$ for this space is provided in Table \ref{tab:basis}, along with the symmetry properties of the basis elements under various index permutations of physical significance. We can express the viscosity tensor as a linear combination of these basis elements:
\begin{equation}
\eta_{ijkl}=\sum_{\alpha=1}^6\lambda_\alpha s^{(\alpha)}_{ijkl}.\label{eq:lambdaexpansion}
\end{equation}

\begin{table*}
	\def\arraystretch{2}
	\setlength{\tabcolsep}{.5em}
	\normalsize\begin{tabular}{|c|c|cccc|c|}\hline
		
		Basis Tensor	&	Components &  $i\leftrightarrow j$ & $k\leftrightarrow l$ &$ij\leftrightarrow kl$ &  P & $s^{(\alpha)}_{ijkl}v_{k,l}$ \\\hline

	$s^{(1)}_{ijkl}$&	$\delta_{ij}\delta_{kl}$ & +& +&  +& +  & $(\boldsymbol{\nabla}\cdot\mathbf{v})\delta_{ij}$\\

		$s^{(2)}_{ijkl}$&	$\delta_{ik}\delta_{j\ell} + \delta_{i\ell}\delta_{jk} -\delta_{ij}\delta_{kl}$& +& +& +& +  & $2\mathbf{\mathring{u}}$\\

	$s^{(3)}_{ijkl}$	&$\epsilon_{ij}\epsilon_{kl}$  & -& -	& +& + & $-2\omega\epsilon_{ij}$\\

		$s^{(4)}_{ijkl}$&	$\epsilon_{ik}\delta_{j\ell}+\epsilon_{j\ell}\delta_{ik}$  & +& +&  -& - & $(\boldsymbol{\sigma}_z\otimes\boldsymbol{\sigma}_x-\boldsymbol{\sigma}_x\otimes\boldsymbol{\sigma}_z):\mathbf{\mathring{u}}$\\

	$s^{(5)}_{ijkl}$	&$\epsilon_{ik}\delta_{j\ell}-\epsilon_{j\ell}\delta_{ik}+\epsilon_{ij}\delta_{k\ell}+\epsilon_{k\ell}\delta_{ij}$ & -	& +& N/A& - & $(\boldsymbol{\nabla}\cdot\mathbf{v})\epsilon_{ij}$ \\

		$s^{(6)}_{ijkl}$	& $\epsilon_{ik}\delta_{j\ell}-\epsilon_{j\ell}\delta_{ik}-\epsilon_{ij}\delta_{k\ell}-\epsilon_{k\ell}\delta_{ij}$ & +& -& N/A& - & $4\omega\delta_{ij}$
		
		\\\hline
	\end{tabular}
	\caption{The tensors $s^{(\alpha)}_{ijkl}$ form a basis for the isotropic rank four tensors in two dimensions, and are orthogonal with respect to the inner product $A_{ijkl}B_{ijkl}$. This basis has been chosen to be an eigenbasis for the index permutations $i\leftrightarrow j$ and $k\leftrightarrow l$, corresponding to the symmetry and objectivity of the stress tensor, respectively. It is also an eigenbasis for the mirror transformation $x_1\mapsto -x_1$, $x_2\mapsto x_2$, also known as the parity transformation (P). Four of these basis tensors are also eigenvectors of the index permutation $i\leftrightarrow k$ and $j\leftrightarrow l$, and we also indicate the parity of the basis tensors under this transformation. In the last column, we provide the component of the stress $T_{ij}=\eta_{ijkl}v_{k,l}$ due to each basis element of the viscosity. The symmetric traceless velocity gradient is defined as $\mathring{u}_{ij}=\frac{1}{2}(v_{i,j}+v_{j,i}-v_{k,k}\delta_{ij})$, and the vorticity as $\omega=-\frac{1}{2}\epsilon_{ij}v_{i,j}$. We also use the matrices $\boldsymbol{\sigma}_z=\left[1,0;0,-1\right]$ and $\boldsymbol{\sigma}_x=\left[0,1;1,0\right]$, which are basis elements of the pure shear modes of the velocity gradient that transform into each other under rotation. The tensor $\boldsymbol{\sigma}_z\otimes\boldsymbol{\sigma}_x$ maps a pure shear mode of the velocity gradient to a rotated pure shear mode of the stress. A complete eigenbasis $\boldsymbol{e}^{(\beta)}$ for the permutation $ij\leftrightarrow kl$ is presented in Appendix I.}
	\label{tab:basis}
\end{table*}

In Table \ref{tab:basis}, we also provide the components $s^{(\alpha)}_{ijkl}v_{k,l}$ of the stress tensor due to each of the basis tensors, elucidating the physical meaning of each coefficient. The bulk viscosity $\lambda_1$ and shear viscosity $\lambda_2$ resist compression and shearing as in a typical Newtonian fluid. The rotational viscosity $\lambda_3$ resists rotation, corresponding to the appearance of a torque in response to non-vanishing vorticity, breaking both symmetry and objectivity of the stress tensor. All three of these components of the viscosity are even under mirror symmetry, implying that they may arise in non-chiral systems.

The other three components of the viscosity are odd under mirror symmetry, and thus should be expected to vanish in non-chiral systems. The odd viscosity $\lambda_4$, corresponding to a term that violates the permutation symmetry $\eta_{ijkl}=\eta_{klij}$, responds to pure shear along one axis with pure shear stress along an axis rotated by $\pi/4$. Equivalently, it responds to simple shear along one axis with pressure or tension along the orthogonal axis, depending on the sign of the shear. An interesting feature of this component of the viscosity is that it is  non-dissipative in the sense that it does not contribute to the stress power $T_{ij}v_{i,j}$. This term satisfies both objectivity and symmetry of the stress tensor, so that it is compatible with conservation of angular momentum even in the absence of internal spin.

Finally, the component $\lambda_5$ responds to compression with torque, breaking symmetry of the stress, while the component $\lambda_6$ responds to vorticity with isotropic pressure, breaking objectivity. The corresponding basis tensors $\mathbf{s}^{(5)}$ and $\mathbf{s}^{(6)}$ both violate the symmetry $\eta_{ijkl}=\eta_{klij}$. They span a two-dimensional subspace with one even and one odd direction under this index permutation, so that there are in fact two independent odd components of the viscosity.\\

\noindent\textbf{\textit{Onsager's regression hypothesis and Green-Kubo relations.}} In what follows we derive Green-Kubo formulae relating the viscous coefficients introduced in \eqref{eq:gammas}-\eqref{eq:lambdaexpansion} to the stress-stress time correlation function in a fluctuating steady-state, starting from an assumption on the fluctuations in the spirit of Onsager's regression hypothesis. The philosophy adopted here is to suppose that the fields $v_i$, $m$, $T_{ij}$, and $C_i$ are fluctuating or stochastic rather than deterministic, but that small fluctuations about a steady state behave, in expectation, in the same manner as the deterministic transport equations would predict. In other words, a viscous fluid is best described in different regimes by either a deterministic or a stochastic theory, and the two must be related in some plausible way. This is the informal content of Onsager's regression hypothesis \cite{Onsager1931a,Onsager1931b, Kubo1957, Kubo1957b}. 

We now provide a more formal presentation of the statement of the regression hypothesis in a general setting, which will be central to the derivation of Green-Kubo relations. Suppose a system is characterized by some set of complex variables $A_i$ and $B_j$, and that the system is described by a deterministic theory that obeys the linear dynamical (or conservation) and constitutive equations
\begin{align}
\frac{dA_i}{dt}=\sum_j M_{ij}B_j,\label{eq:maindyn}\\
B_j=\sum_iS_{ji}A_i,\label{eq:maincon}
\end{align}	
where $M_{ij}$ and $S_{kl}$ are constant coefficients, and $A_i=B_j=0$ is a stable fixed point. These lead to the transport equations 
\begin{align}
\frac{dA_i}{dt}=\sum_{j,k} M_{ij}S_{jk}A_k,\label{eq:maintrans}
\end{align}
where any external perturbation to the variables $A_i$ decays with a  characteristic relaxation time $\tau_\text{r}\approx\frac{1}{M_{ij}S_{jk}}$.

Extending Onsager's regression hypothesis to non-equilibrium steady states \cite{Onsager1931b}, we suppose that spontaneous fluctuations about the steady state decay according to the transport equation \eqref{eq:maintrans} in expectation, in the sense that 
\begin{equation}
\label{eq:transreg}\frac{\brak{A_i(t+\Delta t)}_{t,\mathbf{a}} - a_i}{\Delta t}
=\sum_{j,k} M_{ij}S_{jk}a_k,
\end{equation}
where the subscript indicates that the expectation is taken over the subensemble of trajectories satisfying $A_i(t)=a_i$. 
In \eqref{eq:transreg}, $\Delta t$ is chosen to be small in comparison with the macroscopic relaxation time $\tau_{\text{r}}$, but sufficiently large compared to the microscopic or molecular time scales \cite{Onsager1931b, Kubo1957b}. This is the mathematical statement of the regression hypothesis.

Following Kubo-Yokota-Nakajima \cite{Kubo1957b}, we may derive from the regression \eqref{eq:transreg} (see Appendix III) the generalized Green-Kubo relations
\begin{align}
M_{ij}S_{jk}\brak{A_k(0){A}^*_r(0)}&= -\int_{0}^\infty\brak{\dot{A}_i(t) \dot{A}_r^*(0)}dt, \label{eq:GK-A} \\
& = {-}M_{ij}{M}^*_{rk}\int_{0}^\infty\brak{B_j(t) {B}^*_k(0)}dt,
\end{align} 
where the averages are taken over the steady state ensemble of trajectories, $(\cdot)^{*}$ denotes complex conjugation, and repeated indices are summed. Deriving \eqref{eq:GK-A} requires an important condition on the separation of time scales:
\begin{align}
    \tau_{\text{corr}} \ll \Delta t \ll \frac{1}{M_{ij}S_{jk}},
\end{align}
where $\tau_{\text{corr}}$ is the time scale associated with the decay of the correlation functions $\brak{\dot{A}_j(t) \dot{A}^*_r(0)}$.

\noindent\textbf{\textit{Green-Kubo relations for viscous fluids.}} For our system of non-equilibrium fluids, the role of the variables $A_i$ and $B_j$ will be played by the large wavelength components of the fluctuations of the fields $v_i$, $m$, $T_{ij}$, and $C_i$ about the steady state with $v_i=0$ and $m=\textit{constant}$. The evolution of these components is governed by the linearized Fourier forms of the linear and angular momentum balance equations (see Appendix II). Invoking a regression hypothesis on these variables in the spirit of \eqref{eq:transreg} and examining the large wavelength limit of the fluctuations yields the following Green-Kubo relations for the transport coefficients (see Appendix V for detailed derivations):
\begin{align}
\gamma_1&=\frac{1}{2\rho_0\nu}\delta_{ij}\epsilon_{kl} \mathcal{T}^{ijkl},\label{eq:GK1main}\\
\gamma_2&=\frac{1}{2\rho_0\nu}\epsilon_{ij}\epsilon_{kl}\mathcal{T}^{ijkl},\label{eq:GK2main}\\
\lambda_1+2\lambda_2+\lambda_3- \frac{\gamma_1\pi}{2\mu}+\frac{\gamma_2\tau}{2\mu}&=\frac{1}{2\rho_0\mu}\delta_{ik}\delta_{jl} \mathcal{T}^{ijkl},\label{eq:GK3main}\\
\lambda_4+\lambda_5+\lambda_6- \frac{\gamma_1\tau}{4\mu}-\frac{\gamma_2\pi}{4\mu}&=\frac{1}{4\rho_0\mu}\epsilon_{ik}\delta_{jl} \mathcal{T}^{ijkl},\label{eq:GK4main}\\
\lambda_5-\frac{\gamma_2 \pi}{4\mu}&=\frac{1}{8\rho_0\mu} \epsilon_{ij}\delta_{kl}\mathcal{T}^{ijkl},\label{eq:GK5main}\\
\lambda_3 +\frac{\gamma_2 \tau}{2\mu}&=\frac{1}{4\rho_0\mu}  \epsilon_{ij}\epsilon_{kl}\mathcal{T}^{ijkl}\label{eq:GK6main},
\end{align}
where $\mathcal{T}^{ijkl}$ is the time-integrated stress-stress correlator
\begin{equation}
\mathcal{T}^{ijkl}=\frac{1}{L^4}\int_{0}^\infty dt \int d^2\mathbf{x}\,d^2\mathbf{y} \brak{\delta T_{ij}(\mathbf{x},t)\delta T_{kl}(\mathbf{y},0)},\label{eq:stress-stress-correlator}
\end{equation}
and $\mu$, $\nu$, $\tau$, and $\pi$ are the steady-state correlation functions defined by
\begin{align}
\mu\delta_{ij}&= \frac{1}{L^4}\int \brak{\delta v^i(\mathbf{x})\delta v^j(\mathbf{y})}d^2\mathbf{x}\,d^2\mathbf{y}\\
\pi&= \frac{1}{L^4}\int (y^i-x^i)\brak{\delta v^i(\mathbf{x})\delta m(\mathbf{y})}d^2\mathbf{x}\,d^2\mathbf{y}\\
\tau&= \frac{1}{L^4}\int \epsilon_{kr}(y^r-x^r)\brak{\delta m(\mathbf{x})\delta v^k(\mathbf{y})}d^2\mathbf{x}\,d^2\mathbf{y}\\
\nu&= \frac{1}{L^4}\int \brak{\delta m(\mathbf{x})\delta m(\mathbf{y})}d^2\mathbf{x}\,d^2\mathbf{y}.\label{eq:nu}
\end{align}
In \eqref{eq:GK1main}-\eqref{eq:nu}, $\delta a$ indicates the fluctuation about the steady-state value of $a$. 

The constants $\mu$ and $\nu$ provide an estimate of the effective kinetic temperature in the steady state and by the equipartition theorem are proportional to the Boltzmann temperature in the special case of equilibrium systems. The constant $\tau$ measures the correlation of the internal spin with the fluid vorticity, in other words the correlation between the internal and external angular momentum density fields. The constant $\pi$ measures the correlation of the internal spin with the fluctuating divergence of the velocity field.

Several features of the Green-Kubo relations \eqref{eq:GK1main}-\eqref{eq:GK6main} are noteworthy. In the absence of internal spin (or the absence of a mechanism for coupling internal spin to the velocity field), $\gamma_1=\gamma_2=0$ by assumption and $\lambda_3=\lambda_5=0$ by conservation of angular momentum. Then we are left with the Green-Kubo relations
\begin{align}
\lambda_1+2\lambda_2&=\frac{1}{2\rho_0\mu}\delta_{ik}\delta_{jl} \mathcal{T}^{ijkl},\\
\lambda_4+\lambda_6&=\frac{1}{4\rho_0\mu}\epsilon_{ik}\delta_{jl} \mathcal{T}^{ijkl}.
\end{align}
%\ch{I might be missing something \dots Does the choice of tensor basis prevent us from separately defining $\lambda_1$ and $\lambda_2$? But we know from the standard derivation which of the four components of $\delta_{ik} \delta_{jl} \mathcal{T}^{ijkl}$ go into $\lambda_1$ and which into $\lambda_2$ \dots}
%\km{Cory's comment!}
If we demand that the stress tensor be objective, then $\lambda_6=0$ and we are left with a Green-Kubo relation for the odd viscosity $\lambda_4$. Given the form \eqref{eq:stress-stress-correlator} of the integrated stress-stress correlation function $\mathcal{T}^{ijkl}$, it is clear that only the component of the stress autocorrelation function that is odd under time reversal survives contraction with $\epsilon_{ik}$ as appears in \eqref{eq:GK4main}. Therefore, non-vanishing odd viscosity $\lambda_4\neq 0$ requires breaking time reversal symmetry at the level of the steady-state stress fluctuations for fluids without internal spin.

Now allowing for coupling of internal spin to the fluid velocity, we may observe from \eqref{eq:GK2main} and \eqref{eq:GK6main} that 
\begin{equation}
2\lambda_3=\left(\frac{\nu-\tau}{\mu}\right)\gamma_2.
\end{equation}
In an equilibrium system, $\nu=\mu$ by equipartition and there exist no correlations between internal spin and vorticity, so that $\tau = 0$. Then $\gamma_2=2\lambda_3$, so that there is a single parameter characterizing the response of the stress to both the spin $m$ and the vorticity $\omega$. This feature is assumed in many previous works on out-of-equilibrium active systems \cite{van2016spatiotemporal, Ban17, souslov2019topological, furthauer2012active, furthauer2013active}. It should be noted that such active systems may break equipartition in the steady state so that in general $\nu-\tau\neq \mu$, which leads to decoupling of the two rotational viscosity coefficients coupling the vorticity and internal spin, and therefore this assumption must be revisited.

Finally, we note that in a system with internal spin that obeys time reversal symmetry at the level of the stress correlations, the Green-Kubo relation \eqref{eq:GK4main} involving the odd viscosity reduces to
\begin{align}
\lambda_4+\lambda_5+\lambda_6&= \frac{\gamma_1\tau}{4\mu}+\frac{\gamma_2\pi}{4\mu}.
\end{align}
Thus $\lambda_4$ need not necessarily vanish. Therefore, it is possible that there are systems that do not break time reversal symmetry at the level of stress correlations, yet do exhibit odd viscosity due to a coupling of internal spin to fluid velocity. This possibility merits future consideration. \\

\noindent\textbf{\textit{Conclusion.}}
In this work, we have made progress towards the goal of understanding transport phenomena in systems that break time reversal symmetry. By deriving Green-Kubo formulae via an Onsager regression hypothesis, we have put on stronger footing the claim that in systems without internal spin, non-vanishing odd viscosity requires breaking time reversal symmetry at the level of stress-stress correlations. However, in systems with internal spin, we cannot rule out the possibility of non-vanishing odd viscosity even when this symmetry is preserved. Furthermore, we have demonstrated that breaking of equipartition leads to modification of the coupling between internal spin and fluid vorticity. Future work will attempt to show how these phenomena emerge in particular microscopic models of active systems. The Green-Kubo formulae we derive provide a route to estimating the transport coefficients in molecular simulations \cite{jones2012adaptive,gao2017transport} involving microscopic models and analyzing the nature of non-equilibrium steady states. It will also be interesting to consider the effect of time reversal symmetry breaking on other transport processes involving truly thermodynamic quantities, such as multi-component diffusion and heat transfer.\\

\noindent\textbf{\textit{Acknowledgements.}} The authors thank Cory Hargus for a careful reading of the manuscript and stimulating discussions. They also thank Katie Klymko for insightful discussions. J.M.E.
was supported by the Department of Defense (DoD) through the
National Defense Science and Engineering Graduate Fellowship
(NDSEG) Program. K.K.M was supported by Director, Office of Science, Office of Basic Energy Sciences, of the U.S. Department of Energy under contract No. DEAC02-05CH11231.

%\bibliography{AllBibs}


\onecolumngrid

\def\thesection{\Roman{section}}
\def\thesubsection{\Roman{section}.\Roman{subsection}}
\def\thesubsubsection{\Roman{section}.\Roman{subsection}.\Roman{subsubsection}}
\onecolumngrid
\newpage
\begin{center}
\textbf{Supplementary Information for} \\
\textbf{Time reversal symmetry breaking in two-dimensional\\ non-equilibrium viscous fluids}    \\
\vspace{0.05in}
Jeffrey M. Epstein and Kranthi K. Mandadapu 
\end{center}

In this Supplementary Information, we provide the detailed derivations required to arrive at the results referenced in the main text. In Section I, we prove for convenience a well-known and useful representation theorem for isotropic tensors of any rank in any dimension, namely that they may be represented in terms of the Kronecker and Levi-Civita tensors alone. This allows us to generate a complete basis for the viscosity tensors. In Section II, we describe the balance equations for fluids with internal spin as well as the constitutive equations relating the stress and couple stress to the velocity and spin fields, and describe the assumptions we make about the regime in which we describe the fluid. In section III, we introduce Onsager's regression hypothesis and provide a derivation of a generalized Green-Kubo equation from a set of assumptions on the decay of fluctuations of observables about a non-equilibrium steady state. In Section IV, we discuss the consequences of isotropy for the forms of various correlation functions. In Section V, we combine these tools to derive Green-Kubo equations for the various components of the viscosity from a regression hypothesis on the steady-state fluctuations of the viscous fluid.

\subsection{I. Representation Theorem for Rank-$m$ Isotropic Tensors in Dimension $n$}
It is well known that the pair $(\boldsymbol{\delta},\boldsymbol{\epsilon})$, where $\boldsymbol{\delta}$ is the rank two Kronecker tensor and $\boldsymbol{\epsilon}$ is the rank $n$ Levi-Civita or fully antisymmetric tensor, generate all isotropic tensors on dimension $n$. Recall that the components of these tensors are
\begin{align}
\delta_{ij}&=\begin{cases}
1 & i=j\\
0 & i\neq j,
\end{cases}
\end{align}
and
\begin{align}
\epsilon_{i_1\ldots i_n}&=\begin{cases}
1 & i_1\ldots i_n\text{ even permutation of 1\ldots n}\\
-1 & i_1\ldots i_n\text{ odd permutation of 1\ldots n}\\
0 & \text{otherwise}.
\end{cases}
\end{align}

The sense in which these are generators is made clearer by the graphical calculus frequently used for tensor manipulation in the tensor network community, see for example \cite{bridgeman2017hand} for an introduction to this streamlined language. Briefly, rank $m$ tensors are represented by boxes with $m$ ``legs", each representing one of the indices. Connecting two legs corresponds to identifying the corresponding indices and performing a summation over all possible values of the index, so that for example the scalar product of two rank four tensors is represented pictorially by the diagram in Fig. 1a. The absence of ``free legs" indicates that this is indeed a scalar. In this graphical language, the representation theorem is simple to state: a basis for the isotropic tensors of order $m$ in $n$ dimensions may be obtained by contracting $m$ indexed legs with copies of the $\boldsymbol{\delta}$ and $\boldsymbol{\epsilon}$ tensors in all possible ways, which we may visualize as in Fig. 1b. For example, in two dimensions we can construct the isotropic rank four tensors drawn in Fig. 1c, which are written in component notation as $\delta_{ij}\epsilon_{kl}$, $\delta_{il}\delta_{kj}$, and $\delta_{ik}\delta_{jl}$. We note the curious circumstance, due to the fact that viscosity is rank four and the Levi-Civita tensor is rank $n$, that only in dimensions two and four is there a possibility of a component of the viscosity manifesting breaking of mirror symmetry.

Proofs of the fact that $\boldsymbol{\delta}$ and $\boldsymbol{\epsilon}$ tensors generate all isotropic tensors in dimension $n$ appear in both \cite{weyl1946classical} and \cite{jeffreys1973isotropic}, but the former reference proves a more general theorem using more powerful machinery, while the latter is (for the authors) somewhat difficult to follow. We provide here for convenience a compact proof. The essential idea is simply that the two geometrical quantities that are preserved by rotations of $n$-dimensional Euclidean space are the inner product between pairs of vectors and the signed volume of the parallelepiped spanned by $n$ vectors. These correspond to the tensors $\boldsymbol{\delta}$ and $\boldsymbol{\epsilon}$. The only point that must be verified is that nothing else is preserved. We approach the proof in three steps.\\

\begin{figure}[t!]
    \centering
    \includegraphics[width=.6\textwidth]{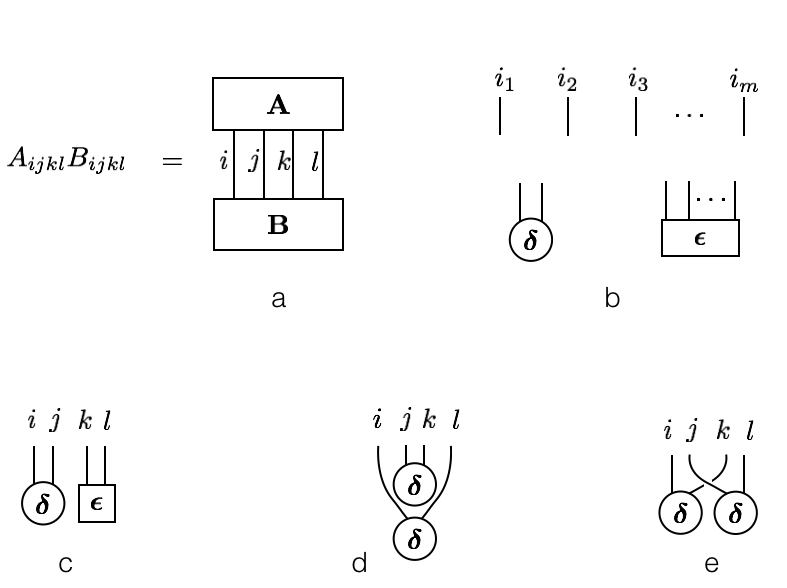}
    \caption{In (a) the scalar product of two fourth-order tensors is represented graphically. In (b) the components for the construction of a basis for all $m^\text{th}$-order isotropic tensors in some dimension $n$ are presented. The idea is that the indexed legs should be connected with copies of the two generator tensors $\boldsymbol{\delta}$ and $\boldsymbol{\epsilon}$. The set of all possible such diagrams spans the space of isotropic tensors. The tensor $\delta_{ij}\epsilon_{kl}$ is represented graphically in (c), $\delta_{il}\delta_{jk}$ in (d), and $\delta_{ik}\delta_{jl}$ in (e). Note that the ``crossing" of the two legs in the third diagram is purely for visual convenience. There is no difference between ``left over right" and ``right over left".}
    \label{fig:graphical}
\end{figure}

\textit{Let $v_1,\ldots,v_m$ and $u_1,\ldots,u_m$ be ordered $m$-tuples of vectors in $\mathbb{R}^n$. Let $\brak{w_1,w_2}$ denote the inner product of vectors $w_1$ and $w_2$ and $\left[w_1,w_2,\ldots,w_n\right]$ denote the determinant of the matrix whose $i^\text{th}$ column is $w_i$. Then the following hold:}
	\begin{enumerate}
		\item  \textit{If $\brak{v_i,v_j}=\brak{u_i,u_j}$ for all $i,j$, then for some $Q\in O(n)$ we have $u_i=Qv_i$ for all $i$.}
		
		\item  \textit{If moreover $\left[v_{i_1},v_{i_2},\ldots,v_{i_n}\right]=\left[u_{i_1},u_{i_2},\ldots,u_{i_n}\right]$ for all $n$-tuples of index assignments, then for some $Q\in SO(n)$ we have $u_i=Qv_i$ for all $i$.}
	\end{enumerate}
 We can establish this result by making use of the notion of Gram-Schmidt orthogonalization. Consider forming the $n\times n$ matrix $E_0$ as follows:
 \begin{enumerate}
     \item Begin by setting $E_0$ equal to the $n\times 1$ column vector $v_1$.
     
     \item For $i$ from 2 to $m$, check whether $v_i$ is in the column space of $E_0$, and if it is not, append $v_i$ as the right-most column of $E_0$.
     
     \item If $E_0$ has fewer than $n$ columns, append columns to the right such that the column space of $E_0$ is $\mathbb{R}^n$.
 \end{enumerate} 
 Now we may define an orthogonal matrix $E\in O(n)$ by performing Gram-Schmidt orthogonalization on the columns of $E_0$. Let the columns of $E$ be denoted by $e_\alpha$, $\alpha=1,2,\ldots n$. The $n\times m$ matrix $V$ with columns $v_i$ may then be expressed as $V=EM$, where $M_{\alpha j}=\brak{e_\alpha,v_j}$.\\
 \\
 We may proceed analogously with the $u_i$ to define matrices $\tilde{E}$, $\tilde{M}$, and $U$ such that $U=\tilde{E}\tilde{M}$. Because the inner products $\brak{e_\alpha,v_j}$ and $\brak{\tilde{e}_\alpha,u_j}$ depend only on pairwise inner products of the $v_i$ and the $u_i$, respectively, we may conclude that $M=\tilde{M}$. Then $U=\tilde{E}M=\tilde{E}E^TEM=\tilde{E}E^TV\equiv QV$, so that $u_i=\tilde{E}E^Tv_i$ for all $i$. Because $E$ and $\tilde{E}$ (and their transposes) are orthogonal, so is $Q=\tilde{E}E^T$, establishing the first point above.\\
 \\
In order to establish the second point, we need to consider the determinant of $Q=\tilde{E}E^T$. Because orthogonal matrices have determinant $\pm 1$ and the Gram-Schmidt procedure uses matrix column operations that do not change the sign of the determinant, we have \begin{align}
    \det E&=\text{sgn}\det E_0    \\
      \det \tilde{E}&=\text{sgn}\det \tilde{E}_0.
\end{align}
Then $\tilde{E}E^T\in SO(n)$ if and only if $\det E=\det\tilde{E}$.\\
\\
If the $v_i$ and $u_i$ independently span $\mathbb{R}^n$, then all of the columns of $E_0$ and $\tilde{E}_0$ are members of these sets of vectors. Then the determinants are equal if and only if $\left[v_{i_1},v_{i_2},\ldots,v_{i_n}\right]=\left[u_{i_1},u_{i_2},\ldots,u_{i_n}\right]$ for all choices $i_1,\ldots, i_n$. If the $v_i$ and $u_i$ do not span $\mathbb{R}^n$, we are free to choose the extra basis vectors, so may arrange to have $\det(\tilde{E}E^T)=1$. In this situation, $\left[v_{i_1},v_{i_2},\ldots,v_{i_n}\right]=\left[u_{i_1},u_{i_2},\ldots,u_{i_n}\right]=0$. This establishes the second point above.\\
\\
This fact allows us to prove the following:\\
\\
\textit{
	Suppose that $f$ is a function of $m$-tuples of vectors in $\mathbb{R}^n$ that is invariant under the orthogonal group $O(n)$ in the sense that $f(Qv_1,\ldots,Qv_n)=f(v_1,\ldots,v_n)$ for any $Q\in O(n)$. Then $f$ can be expressed as a function of the pairwise inner products $\brak{v_i,v_j}$. If $f$ is only required to be invariant under the special orthogonal group $SO(n)$, then $f$ may be expressed as a function of the pairwise inner products $\brak{v_i,v_j}$ and the determinants $\left[v_{i_n},\ldots,v_{i_n}\right]$.}\\
\\
	Let $f$ be invariant under $O(n)$. Suppose that $\brak{v_i,v_j}=\brak{u_i,u_j}$ for all $i,j$ and that $f(u_1,\ldots, u_m)\neq f(v_1,\ldots v_m)$. By the result given above, the equality of pairwise inner products implies that there is some $Q\in O(n)$ such that $u_i=Qv_i$ for all $i$. Then $f(Qv_1,\ldots, Qv_m)\neq f(v_1,\ldots,v_m)$. But this contradicts the assumption that $f$ is invariant under $O(n)$. Therefore, $f$ must be completely determined by the set of pairwise inner products $\brak{v_i,v_j}$.\\
	\\
	Now let $f$ be invariant under $SO(n)$ but not necessarily under all of $O(n)$. Suppose that $\brak{v_i,v_j}=\brak{u_i,u_j}$ and $\left[v_{i_n},\ldots,v_{i_n}\right]=\left[u_{i_n},\ldots,u_{i_n}\right]$ for all choices of index assignments $f(u_1,\ldots, u_m)\neq f(v_1,\ldots v_m)$. By the above result, there is some $Q\in SO(n)$ such that $u_i=Qv_i$ for all $i$, so $f(Qv_1,\ldots, Qv_m)\neq f(v_1,\ldots,v_m)$. Again, this leads to a contradiction, so $f$ must be determined by the set of inner products and determinants.\\
	\\
	Now we are equipped to prove the promised representation theorem:

\begin{theorem}
	Let $T$ be an order $m$ tensor on dimension $n$, in other words a multilinear map
	\begin{equation}
	T:\underbrace{\mathbb{R}^n\otimes \mathbb{R}^n\otimes\cdots\otimes \mathbb{R}^n}_{m\text{ times}}\rightarrow\mathbb{R}.
	\end{equation}
	Suppose that $T$ is invariant under the special orthogonal group $SO(n)$ of proper rotations in the sense that for any $Q\in SO(n)$ and any $u_1,u_2,\ldots,u_m\in\mathbb{R}^n$, we have
	\begin{equation}
	T(Qu_1\otimes Qu_2\otimes\cdots\otimes Qu_m)=T(u_1\otimes u_2\otimes\cdots\otimes u_m).
	\end{equation}
	Then $T$ may be expressed as a linear combination of products of order 2 Kronecker tensors $\delta$ and at most one order $n$ alternating/Levi-Civita tensors $\epsilon$ acting on disjoint sets of tensor factors (indices). If $T$ is invariant under the entire orthogonal group $O(n)$, then it is expressible only in terms of Kronecker tensors.
\end{theorem}
By invariance under $SO(n)$, we can conclude using the previous result that $T(u_1\otimes u_2\otimes\cdots\otimes u_m)$ is determined completely the pairwise inner products $\brak{u_i,u_j}$ and volumes $\left[u_{i_1},\ldots,u_{i_n}\right]$. For $T$ to be linear in each of its arguments, this function must be a linear combination of products of inner products and volume forms in which each argument $u_i$ appears exactly once. The inner product is given by $\brak{u,v}=\delta(u\otimes v)$ with $\delta$ the Kronecker tensor, and the volume form is given by $\left[u,v,\ldots, w\right]=\epsilon(u\otimes v\otimes\cdots\otimes w)$ with $\epsilon$ the alternating tensor. It is easy to see that the alternating tensor itself is odd under reflections so that a product of two alternating tensors on disjoint sets of indices is even. Then such products may be expressed solely in terms of Kronecker tensors. This establishes the first part of the theorem. If $T$ is required to be invariant under $O(n)$, it can't contain any terms with an odd number of $\epsilon$ tensors, as these are odd under parity-inverting transformations. This establishes the second, and the representation theorem is proven.\\

This representation theorem provides a method for generating and studying the possible isotropic viscosity tensors, as we may use it to construct orthonormal bases for the full space of isotropic rank four tensors in two dimensions that diagonalize any symmetry of interest. In two dimensions, the space of isotropic rank four tensors is six-dimensional. Two different orthogonal bases (eigenbases for different sets of symmetries) for this space are presented in Tables \ref{tab1} and \ref{tab2}.

\begin{table}[t!]
	\def\arraystretch{2}
	\setlength{\tabcolsep}{.5em}
	\normalsize\begin{tabular}{|c|c|ccc|}\hline
		
		Basis Tensor	&	Components  & $j\leftrightarrow l$ &$ij\leftrightarrow kl$ &  P \\\hline
		
		$\mathbf{e}^{(1)}$& $\delta_{ij}\delta_{kl}-\epsilon_{ij}\epsilon_{kl}$&  +& +& + \\
		
	$\mathbf{e}^{(2)}$&$\epsilon_{ik}\epsilon_{jl}$ 	& -& +& + \\
		
		$\mathbf{e}^{(3)}$&	$\delta_{ik}\delta_{jl}$	& +& +& + \\

		$\mathbf{e}^{(4)}$&	$\epsilon_{ik}\delta_{jl}$ & +& -& - \\

		$\mathbf{e}^{(5)}$&$\epsilon_{ij}\delta_{kl}+\epsilon_{kl}\delta_{ij}$ 	& +& +& - \\

	$\mathbf{e}^{(6)}$&	$\epsilon_{jl}\delta_{ik}$& -&-& -	
		
		\\\hline
	\end{tabular}
	\caption{Basis for the isotropic rank four tensors in two dimensions in which the index permutations $j\leftrightarrow l$ and $i\leftrightarrow k$, $j\leftrightarrow l$ are diagonal. Components of the viscosity odd under the former do not contribute to the momentum balance, while components odd under the latter contribute to the odd viscosity. The mirror transformation $(x_1,x_2)\mapsto(-x_1,x_2)$ is also diagonal in this basis. Note that the $+(-)$ indicates that the basis tensor is even (odd) under the indicated transformation. The basis tensors are orthogonal (but not normalized) with respect to the inner product $A_{ijkl}B_{ijkl}$.}\label{tab1}
\end{table}

\begin{table}[t!]
	\def\arraystretch{2}
	\setlength{\tabcolsep}{.5em}
	\normalsize\begin{tabular}{|c|c|cccc|}\hline
		
		Basis Tensor	&	Components &  $i\leftrightarrow j$ & $k\leftrightarrow l$ &$ij\leftrightarrow kl$ &  P \\\hline

	$\mathbf{s}^{(1)}$&	$\delta_{ij}\delta_{kl}$ & +& +&  +& +  \\

		$\mathbf{s}^{(2)}$&	$\delta_{ik}\delta_{j\ell}-\epsilon_{ik}\epsilon_{jl}$& +& +& +& +  \\

	$\mathbf{s}^{(3)}$	&$\epsilon_{ij}\epsilon_{kl}$  & -& -	& +& + \\

		$\mathbf{s}^{(4)}$&	$\epsilon_{ik}\delta_{j\ell}+\epsilon_{j\ell}\delta_{ik}$  & +& +&  -& - \\

	$\mathbf{s}^{(5)}$	&$\epsilon_{ik}\delta_{j\ell}-\epsilon_{j\ell}\delta_{ik}+\epsilon_{ij}\delta_{k\ell}+\epsilon_{k\ell}\delta_{ij}$ & -	& +& N/A& - \\

		$\mathbf{s}^{(6)}$	&$\epsilon_{ik}\delta_{j\ell}-\epsilon_{j\ell}\delta_{ik}-\epsilon_{ij}\delta_{k\ell}-\epsilon_{k\ell}\delta_{ij}$& +& -& N/A& - 
		
		\\\hline
	\end{tabular}
	\caption{Basis for the isotropic rank four tensors in two dimensions in which the index permutation $i\leftrightarrow j$, the permutation $k\leftrightarrow l$, and again the mirror transformation are diagonal. All but two of these basis elements are also eigenvectors of the permutation $i\leftrightarrow k$, $j\leftrightarrow l$, so where possible we also indicate the eigenvalue of the basis tensors under this symmetry. It is this basis we use for discussing the Green-Kubo relations presented in the main text.}\label{tab2}
\end{table}

\subsection{II. Balance and Constitutive Equations for Fluctuations About Steady State}
\subsubsection{1. Balance Equations}

In the absence of external sources of linear or angular momentum, the global conservation of these quantities as well as of mass leads to the balance equations
\begin{align}
\dot{\rho}&=-\rho v_{i,i}\\
\rho\dot{v}_i&=T_{ij,j},\\
\rho\dot{m}&=-\epsilon_{ij}T_{ij}+C_{i,i},
\end{align}
where the variables are the mass density $\rho$, velocity $v_i$, internal spin $m$ (scalar in two dimensions), stress tensor $T_{ij}$, couple stress ${C}_{i}$. The dot indicates the material derivative $\partial_t+v^i\partial_i$. Let $\rho_0$, ${v}^i_0$, $m_0$, and so on denote the spatially-uniform values of the fields in a stable steady state. We may express the balance equations in terms of small deviations $\delta\rho$, $\delta{v}^i$, $\delta{m}$, and so on from the steady state:
\begin{align}
\partial_t\delta\rho+(v^j_0+\delta v^j)\delta\rho_{,j}&=-(\rho_0+\delta\rho)\delta v_{i,i},\\
(\rho_0+\delta\rho)\partial_t\delta v_i+(\rho_0+\delta\rho)(v^j_0+\delta v^j)\delta v_{i,j}&=\delta T_{ij,j},\\
(\rho_0+\delta\rho)\partial_t\delta m+(\rho_0+\delta\rho)(v^j_0+\delta v^j)\delta m_{,j}&=-\epsilon_{ij}(T_0^{ij}+\delta T_{ij})+\delta C_{i,i}.
\end{align}
Now requiring $v_0=0$, to linear order in the deviations from the steady state, these become
\begin{align}
\partial_t\delta\rho&=-\rho_0\delta v_{i,i},\\
\rho_0\partial_t\delta v_i&=\delta T_{ij,j},\\
\rho_0\partial_t\delta m&=-\epsilon_{ij}\delta T_{ij}+\delta C_{i,i},
\end{align}
where we have used the fact that $\epsilon_{ij}T_{ij}$ must vanish in a steady state.

Taking the fields to be defined on a square region of side length $L$ with periodic boundary conditions, we may decompose them into Fourier components, so that
\begin{align}
\delta\rho(\mathbf{x},t)&=\sum_\mathbf{k}\rho_\mathbf{k}(t)e^{\im\mathbf{k}\cdot\mathbf{x}},\\
\delta v^i(\mathbf{x},t)&=\sum_\mathbf{k}v^i_\mathbf{k}(t)e^{\im\mathbf{k}\cdot\mathbf{x}},\\
\delta m(\mathbf{x},t)&=\sum_\mathbf{k}m_\mathbf{k}(t)e^{\im\mathbf{k}\cdot\mathbf{x}},\\
\delta T_{ij}(\mathbf{x},t)&=\sum_\mathbf{k} T^{ij}_\mathbf{k}(t)e^{\im\mathbf{k}\cdot\mathbf{x}},\\
\delta C_i(\mathbf{x},t)&=\sum_\mathbf{k}C^i_\mathbf{k}(t)e^{\im\mathbf{k}\cdot\mathbf{x}},
\end{align}
with $\mathbf{k}$ taking discrete values. In terms of the Fourier variables, the linearized balance equations then take the forms

\begin{align}
\dot{\rho}_\mathbf{k}&=-\im \rho_0k^iv_\mathbf{k}^i,\\
\rho_0\dot{v}_\mathbf{k}^i&=\im k^j T_\mathbf{k}^{ij},\\
\rho_0\dot{m}_\mathbf{k}&=-\epsilon_{ij}T_\mathbf{k}^{ij}+\im k^iC_\mathbf{k}^i.
\end{align}

\subsubsection{2. Constitutive Equations}

We will assume that the fluid is in a regime in which the mass density $\rho$ is approximately constant, and that the variations in the stress $T_{ij}$ and couple stress ${C}_i$ depend only on the velocity gradient $v_{k,l}$, the internal spin $m$, and the spin gradient $m_{,j}$. In principle, the constants relating these quantities may depend on $\rho$. Under these assumptions, the most general linear constitutive relations for  $T_{ij}$ and $C_i$ are
\begin{align}
T_{ij}&=\eta_{ijkl}v_{k,l}+\gamma_{ij} m+\xi_{ijk}m_{,k},\\
C_{i}&=\beta_{ijk} v_{j,k}+\kappa_i m+\alpha_{ij}m_{,j}.
\end{align}
Now imposing isotropy, as mentioned in the main text, we may eliminate several of the terms by using the fact that there do not exist isotropic tensors of ranks one or three in two dimensions. We find
\begin{align}
T_{ij}&=\eta_{ijkl}v_{k,l}+\gamma_{ij} m,\\
C_{i}&=\alpha_{ij}m_{,j}.
\end{align}
Expressing these in the Fourier basis, we see that
\begin{align}
T^{ij}_\mathbf{k}&=\im \eta_{ijkl}k^lv_\mathbf{k}^{k}+\gamma_{ij} m_\mathbf{k},\label{eq:fourier-str}\\
C_\mathbf{k}^{i}&=\im\alpha_{ij}k^jm_\mathbf{k}. \label{eq:fourier-couple}
\end{align}

\subsubsection{3. Matrix Form of Balance Equations and Constitutive Equations}
The Fourier forms of the balance (or conservation) equations may be expressed in matrix form by
\begin{align}
\frac{d}{dt}\left[\begin{tabular}{c}
$v_\mathbf{k}^1$      \\
$v_\mathbf{k}^2$      \\
$m_\mathbf{k}$
\end{tabular}\right]=\frac{1}{\rho_0}\left[\begin{tabular}{cccccc}
 $\im k^1$ & $\im k^2$ & 0 & 0 & 0 & 0 \\
0 & 0 & $\im k^1$ & $\im k^2$ & 0 & 0\\
 0 & $-1$ & $1$ & 0 & $\im k^1$ & $\im k^2$
\end{tabular}\right]\left[\begin{tabular}{c} 
$T_\mathbf{k}^{11}$ \\
$T_\mathbf{k}^{12}$ \\
$T_\mathbf{k}^{21}$ \\
$T_\mathbf{k}^{22}$ \\
$C_\mathbf{k}^{1}$ \\
$C_\mathbf{k}^{2}$ 
\end{tabular}\right]
\end{align}
or more compactly by
\begin{align}
\frac{d}{dt}\left[\begin{tabular}{c}
$v_\mathbf{k}^r$      \\
$m_\mathbf{k}$
\end{tabular}\right]=\frac{1}{\rho_0}\left[\begin{tabular}{cc}
 $\im k^\nu\delta_{\mu r}$  & 0 \\
 $-\epsilon_{\mu\nu}$  & $\im k^\lambda$ 
\end{tabular}\right]\left[\begin{tabular}{c}  
$T_\mathbf{k}^{\mu\nu}$ \\
$C_\mathbf{k}^{\lambda}$ \\ 
\end{tabular}\right],
\end{align}
where, for convenience, Latin indices have been used to label components of the configuration variables $\mathbf{v}_\mathbf{k}$ and $m_\mathbf{k}$, and Greek indices to label components of the generalized fluxes $\mathbf{T}_\mathbf{k}$ and $\mathbf{C}_\mathbf{k}$. 

The Fourier forms of the constitutive relations \eqref{eq:fourier-str} and \eqref{eq:fourier-couple} may also be cast in matrix form:
\begin{align}
\left[\begin{tabular}{c} 
$T_\mathbf{k}^{11}$ \\
$T_\mathbf{k}^{12}$ \\
$T_\mathbf{k}^{21}$ \\
$T_\mathbf{k}^{22}$ \\
$C_\mathbf{k}^{1}$ \\
$C_\mathbf{k}^{2}$ 
\end{tabular}\right]&=\left[\begin{tabular}{ccc}
  $\im\eta_{111j}k^j$ & $\im\eta_{112j}k^j$& $\gamma_{11}$\\
  $\im\eta_{121j}k^j$ & $\im\eta_{122j}k^j$ & $\gamma_{12}$\\
   $\im\eta_{211j}k^j$ & $\im\eta_{212j}k^j$ & $\gamma_{21}$\\
  $\im\eta_{221j}k^j$ & $\im\eta_{222j}k^j$& $\gamma_{22}$\\
0 &0& $\im\alpha_{1j}k^j$\\
0 &0& $\im\alpha_{2j}k^j$
\end{tabular}\right]\left[\begin{tabular}{c}
$v_\mathbf{k}^1$      \\
$v_\mathbf{k}^2$      \\
$m_\mathbf{k}$
\end{tabular}\right]
\end{align}
and more compactly using the aforementioned convention with the Latin and Green indices as
\begin{align}
\left[\begin{tabular}{c}  
$T_\mathbf{k}^{\mu\nu}$ \\
$C_\mathbf{k}^{\lambda}$ 
\end{tabular}\right]&=\left[\begin{tabular}{cc}
  $\im\eta_{\mu\nu r\aleph}k^\aleph$ & $\gamma_{\mu\nu}$\\
0 & $\im\alpha_{\lambda \aleph}k^\aleph$
\end{tabular}\right]\left[\begin{tabular}{c}
$v_\mathbf{k}^r$      \\
$m_\mathbf{k}$
\end{tabular}\right].
\end{align}
Making the definitions
\begin{align}
\mathbf{A}&=\left[\begin{tabular}{c}
$v_\mathbf{k}^r$      \\
$m_\mathbf{k}$
\end{tabular}\right],\hspace{30pt}\mathbf{B}=\left[\begin{tabular}{c} 
$T_\mathbf{k}^{\mu\nu}$ \\
$C_\mathbf{k}^{\lambda}$ \\ 
\end{tabular}\right],\hspace{30pt}\mathbf{M}=\frac{1}{\rho_0}\left[\begin{tabular}{ccccccc}
$\im k^\nu\delta_{\mu r}$  & 0 &\\
 $-\epsilon_{\mu\nu}$  & $\im k^\lambda$ 
\end{tabular}\right],\hspace{30pt}\mathbf{S}=\left[\begin{tabular}{ccc}
  $\im\eta_{\mu\nu r\aleph}k^\aleph$ & $\gamma_{\mu\nu}$\\
0 & $\im\alpha_{\lambda \aleph}k^\aleph$
\end{tabular}\right],
\end{align}
we see that the small deviations about the steady state obey the dynamical (or conservation) and constitutive equations
\begin{align}
\frac{d}{dt}\mathbf{A}&=\mathbf{M}\mathbf{B}, \label{eq:app-fluid-conser}\\
\mathbf{B}&=\mathbf{S}\mathbf{A}. \label{eq:app-fluid-consti}
\end{align}

\subsubsection{4. Expansion of Balance and Constitutive Equations in Basis Tensors}
We now express $\boldsymbol{\gamma}$ and $\boldsymbol{\eta}$ in terms of bases for the isotropic tensors of rank two and four, respectively, in two dimensions, as discussed in Appendix I:
\begin{align}
\gamma_{ij}&=\gamma_1\delta_{ij}+\gamma_2\epsilon_{ij}\label{eq:gammacomponentform},\\
\eta_{ijkl}&=\sum_{\alpha=1}^6\beta_{\alpha}e^{(\alpha)}_{ijkl}=\sum_{\alpha=1}^6\lambda_\alpha s_{ijkl}^{(\alpha)}\label{eq:etacomponentform}\\
\nonumber &=\beta_1(\delta_{ij}\delta_{kl}-\epsilon_{ij}\epsilon_{kl})+\beta_2\epsilon_{ik}\epsilon_{jl}+\beta_3\delta_{ik}\delta_{jl}+\beta_4\epsilon_{ik}\delta_{jl}+\beta_5(\epsilon_{ij}\delta_{kl}+\epsilon_{kl}\delta_{ij})+\beta_6\epsilon_{jl}\delta_{ik}\\
\nonumber\\
\nonumber&=\lambda_1\delta_{ij}\delta_{kl}+\lambda_2(\delta_{ik}\delta_{jl}-\epsilon_{ik}\epsilon_{jl})+\lambda_3\epsilon_{ij}\epsilon_{kl}+\lambda_4(\epsilon_{ik}\delta_{jl}+\epsilon_{jl}\delta_{ik})\\
\nonumber&\hspace{20pt}+\lambda_5(\epsilon_{ik}\delta_{jl}-\epsilon_{jl}\delta_{ik}+\epsilon_{ij}\delta_{kl}+\epsilon_{kl}\delta_{ij})+\lambda_6(\epsilon_{ik}\delta_{jl}-\epsilon_{jl}\delta_{ik}-\epsilon_{ij}\delta_{kl}-\epsilon_{kl}\delta_{ij}),
\end{align}
where the $\mathbf{e}^{(\alpha)}$ and $\mathbf{s}^{(\alpha)}$ are the basis tensors given in Tables \ref{tab1} and \ref{tab2}. This leads to the constitutive equations for the stress in the $\mathbf{s}^{(\alpha)}$ basis:
\begin{align}
T_{ij}&=\eta_{ijkl}v_{k,l}+\gamma_1m\delta_{ij}+\gamma_2m\epsilon_{ij}\\
\nonumber\\
&=\lambda_1\sone v_{k,l}+\lambda_2\stwo v_{k,l}+\lambda_3\sthree v_{k,l}+\lambda_4\sfour v_{k,l}\\
\nonumber&\hspace{20pt}+\lambda_5\sfive v_{k,l}+\lambda_6\ssix v_{k,l}\\
\nonumber&\hspace{20pt}+\gamma_1m\delta_{ij}+\gamma_2m\epsilon_{ij}\\
\nonumber\\
&=\lambda_1v_{k,k}\delta_{ij}+\lambda_2 (v_{i,j}-v_{k,k}\delta_{ij}+v_{j,i})-2\lambda_3\omega \epsilon_{ij}+\lambda_4(\epsilon_{ik}v_{k,j}+\epsilon_{jl}v_{i,l})\\
\nonumber&\hspace{20pt}+\lambda_5(\epsilon_{ik}v_{k,j}-\epsilon_{jl}v_{i,l}+\epsilon_{ij}v_{k,k}-2\omega\delta_{ij})+\lambda_6(\epsilon_{ik}v_{k,j}-\epsilon_{jl}v_{i,l}-\epsilon_{ij}v_{k,k}+2\omega\delta_{ij})\\
\nonumber&\hspace{20pt}+\gamma_1m\delta_{ij}+\gamma_2m\epsilon_{ij}\\
\nonumber\\
&=(\lambda_1-\lambda_2)v_{k,k}\delta_{ij}+\lambda_2 (v_{i,j}+v_{j,i})+(\gamma_2m-2\lambda_3\omega) \epsilon_{ij}+\lambda_4(\epsilon_{ik}v_{k,j}+\epsilon_{jl}v_{i,l})\\
\nonumber&\hspace{20pt}+(\lambda_5+\lambda_6)(\epsilon_{ik}v_{k,j}-\epsilon_{jl}v_{i,l})+(\lambda_5-\lambda_6)v_{k,k}\epsilon_{ij}+\left[\gamma_1m-2(\lambda_5-\lambda_6)\omega\right]\delta_{ij}\\
\nonumber\\
&=(\lambda_1-\lambda_2)v_{k,k}\delta_{ij}+\lambda_2 (v_{i,j}+v_{j,i})+(\gamma_2m-2\lambda_3\omega) \epsilon_{ij}+(\lambda_4+\lambda_5+\lambda_6)\epsilon_{ik}v_{k,j}\\
\nonumber&\hspace{20pt}+(\lambda_4-\lambda_5-\lambda_6)\epsilon_{jk}v_{i,k}+(\lambda_5-\lambda_6)v_{k,k}\epsilon_{ij}+\left[\gamma_1m-2(\lambda_5-\lambda_6)\omega\right]\delta_{ij},
\end{align}
\noindent where we have used the definition of the vorticity $\omega=-\frac{1}{2}\epsilon_{ij}v_{i,j}$. Substituting the constitutive equations in the linear momentum balance, we obtain the transport equations
\begin{align}\label{eq:lin-mom-gen}
\rho\dot{\mathbf{v}}&=\lambda_1\boldsymbol{\nabla}(\boldsymbol{\nabla}\cdot\mathbf{v})+\lambda_2 \boldsymbol{\Delta}\mathbf{v}+(\lambda_5-\lambda_6)\boldsymbol{\epsilon}\cdot\boldsymbol{\nabla}(\boldsymbol{\nabla}\cdot\mathbf{v})+(\lambda_4+\lambda_5+\lambda_6)\boldsymbol{\epsilon}\cdot\boldsymbol{\Delta}\mathbf{v}\\
\nonumber&\hspace{20pt}+\boldsymbol{\epsilon}\cdot\boldsymbol{\nabla}(\gamma_2m-2\lambda_3\omega)
+\boldsymbol{\nabla}\left[\gamma_1m-2(\lambda_5-\lambda_6)\omega\right].
\end{align}
In the absence of internal spin, the stress must be symmetric, so that $\lambda_3=\lambda_5=0$, and demanding objectivity of the stress, so that $\lambda_3=\lambda_6=0$, equation \eqref{eq:lin-mom-gen} becomes
\begin{align}
\rho\dot{\mathbf{v}}&=\lambda_1\boldsymbol{\nabla}(\boldsymbol{\nabla}\cdot\mathbf{v})+\lambda_2 \boldsymbol{\Delta}\mathbf{v}+\lambda_4\boldsymbol{\epsilon}\cdot\boldsymbol{\Delta}\mathbf{v},
\end{align}
which contains an additional term corresponding to the odd viscosity $\lambda_4$ compared to a typical Newtonian fluid. Analogous transport equations can be derived with the $\mathbf{e}^{(\alpha)}$ basis.

\subsection{III. Onsager's Regression Hypothesis and Green-Kubo Relations}
The physics of matter at large scales is typically captured by theories with deterministic evolution equations. On the other hand, measurements made at small scales with high precision reveal stochastic behavior. As a consequence, a single physical system may be best described in different regimes by two different theories, one deterministic and one stochastic. Clearly, there must be some relation between these. The regression hypothesis proposed by Onsager and used in his derivation of the reciprocal relations of transport coefficients is one possible such relation \cite{Onsager1931a,Onsager1931b}. Informally, the content of the regression hypothesis is that the dynamical and constitutive equations that yield the transport equations obeyed by the variables in a deterministic theory are also satisfied in expectation by the stochastic theory describing fluctuations of the same system about a steady state, when conditioned on initial conditions.

We can provide a more formal account of the regression hypothesis by considering a system described by some generalized configuration variables $A_i$ whose evolution we are interested in modeling. In a fluid, for example, these may be Fourier modes $\mathbf{v}_\mathbf{k}$ of the velocity field. We suppose that these evolve according to some conservation laws associated with generalized flux variables $B_j$, obeying a relationship
\begin{align}
\frac{dA_i}{dt}&=M_{ij}B_j,\label{eq:dyn_det}
\end{align}	
where the repeated index summation convention is used. In a fluid, the variables $B_j$ will be Fourier modes ${T}^{ij}_\mathbf{k}$ of the stress field. We now assume that the generalized fluxes themselves depend on the configuration variables $A_{i}$ via the constitutive relations
\begin{align}
B_j&=S_{ji}A_i,\label{eq:con_det}
\end{align}	
where $S_{ji}$ are transport coefficients. Together, \eqref{eq:dyn_det} and \eqref{eq:con_det} define a deterministic theory and lead to the macroscopic transport equations 
\begin{align}
    \frac{dA_i}{dt}&=M_{ij}S_{jk}A_k.\label{eq:app_trans}
\end{align}

Let $A_i = B_j = 0$ be a fixed point for the transport equation \eqref{eq:app_trans} corresponding to some stable steady state, possibly non-equilibrium. Any external perturbation to the steady state will decay on a time scale $\tau_{\text{r}}$ given by 
\begin{align}
    \tau_{\text{r}} \approx \frac{1}{M_{ij}S_{jk}}.\label{eq:appendix-macro-relax}
\end{align}

Suppose that we are interested in spontaneous fluctuations arising in the steady state. Let $A_i(t)$ be stochastic fluctuations about the steady state $\brak{A_i} = \brak{B_j} = 0$. Any fluctuations in $A_i$ should decay back to zero. Let $A_i(t) = a_i$ at an initial time $t$. There exist many trajectories of the system that are commensurate with such a choice. Let $\brak{A_i(t+\Delta t)}_{t,\mathbf{a}}$ be an average of the observable $A_i$ at time $t+\Delta t$, where the subscript indicates an average of the ensemble compatible with the choice $A_i(t) = a_i$, and $\Delta t$ is sufficiently small compared to the macroscopic relaxation time $\tau_{\text{r}}$ in \eqref{eq:appendix-macro-relax} but sufficiently large compared to a molecular time scale, which will be defined later.  In this case, assuming that the fluctuations are sufficiently small, one may postulate that the decay of the fluctuations towards the steady state satisfies the Onsager's regression hypothesis \cite{Onsager1931a, Onsager1931b, Kubo1957b}, i.e., the decay of the fluctuations follows the same transport equations \eqref{eq:app_trans} in a finite time difference manner given by 
\begin{align}
\frac{\brak{A_i(t+\Delta t)}_{t,\mathbf{a}} - a_i}{\Delta t} =  M_{ij}S_{jk}a_k.\label{eq:app-reg-trans}
\end{align}
Equation \eqref{eq:app-reg-trans} is the form of the regression hypothesis implemented by Kubo-Yokota-Nakajima in their derivation of the Green-Kubo relations for responses of thermal origin \cite{Kubo1957b}.
Note that the conservation laws/dynamical equations \eqref{eq:dyn_det} are valid for every trajectory, and it is the linear constitutive relations that are satisfied in expectation in \eqref{eq:app-reg-trans}. 
In the Irving-Kirkwood framework \cite{Irving1950,epstein2018statistical}, for instance, the spatially coarse-grained stress field is \textit{defined} by taking the derivative of the coarse-grained velocity field, so that the momentum balance equation is exactly satisfied for every trajectory in the ensemble. It is the constitutive relation that introduces uncertainty, as the coarse-grained velocity and density fields do not completely specify the configuration of the individual particles, which would be required for precise knowledge of the coarse-grained stress field. 

We now turn to examining a consequence of the regression hypothesis, namely the general Green-Kubo relations for the transport coefficients $S_{jk}$, by following the procedure adopted in the original derivation of Green-Kubo relations by Kubo-Yokota-Nakajima \cite{Kubo1957b}. To this end, we multiply \eqref{eq:app-reg-trans} by $a_r^*$ to find 
\begin{align}
    \frac{1}{\Delta t} \Big(\brak{A_i(t+\Delta t)}_{t,\mathbf{a}}a_r^* - a_i a_r^* \Big) =  M_{ij}S_{jk}a_k a_r^*.\label{eq:app-der-1}
\end{align}
where $(\cdot)^*$ indicates a complex conjugate. 
Taking average over the entire ensemble on both sides of \eqref{eq:app-der-1} corresponding to all possible values of $a_k$ yields 
\begin{align}
    \frac{1}{\Delta t} \Big[\brak{A_i(t+\Delta t)A_r^*(t)} - \brak{A_i(t) A_r^*(t)} \Big] =  M_{ij}S_{jk}\brak{A_k(t) A_r^*(t)}, \label{eq:app-der-2}
\end{align}
which then reduces to 
\begin{align}
    \frac{1}{\Delta t} \Big[\brak{A_i(\Delta t)A_r^*(0)} - \brak{A_i(0) A_r^*(0)} \Big] =  M_{ij}S_{jk}\brak{A_k(0) A_r^*(0)}. \label{eq:app-der-3}
\end{align}
Consider the left hand side of \eqref{eq:app-der-3}:
\begin{align}
    \frac{1}{\Delta t} \Big[\brak{A_i(\Delta t)A_r^*(0)} - \brak{A_i(0) A_r^*(0)} \Big] &=  \frac{1}{\Delta t} \Bigg[\brak{\Big(A_i(0) + \int_0^{\Delta t}\dot{A}_i(t') dt'\Big) A_r^*(0)} - \brak{A_i(0) A_r^*(0)} \Bigg] \\
    & = \frac{1}{\Delta t} \Bigg[\int_0^{\Delta t} dt' \brak{\dot{A}_i(t')  A_r^*(0)}  \Bigg],
\end{align}
which reduces \eqref{eq:app-der-3} to 
\begin{align}
     \frac{1}{\Delta t} \Bigg[\int_0^{\Delta t}dt'\brak{\dot{A}_i(t')  A_r^*(0)}\Bigg] =  M_{ij}S_{jk}\brak{A_k(0) A_r^*(0)}. \label{eq:app-der-4}
\end{align}

Consider the following time derivative of correlation function: 
\begin{align}
     \frac{d}{d \tau} \brak{\dot{A}_i(\tau)  A_r^*(0)} &= \frac{d}{d \tau} \brak{\dot{A}_i(0)  A_r^*(-\tau)}, \\
     & = -\brak{\dot{A}_i(0)  \dot{A}_r^*(-\tau)}, \\
     &  = - \brak{\dot{A}_i(\tau)  \dot{A}_r^*(0)}, \label{eq:app-der-5}
\end{align}
which yields 
\begin{align}
      \brak{\dot{A}_i(\tau)  A_r^*(0)} &= - \int_0 ^{\tau} d t'' \brak{\dot{A}_i(t'')  \dot{A}_r^*(0)} + \brak{\dot{A}_i(0)  A_r^*(0)}. \label{eq:app-der-6}
\end{align}
At this stage, we assume that the steady-state time correlation of observables $A_i(t)$ and $A_r(0)$ reaches an extremum at $t=0$, so that 
\begin{align}
    \brak{\dot{A}_i(0)  A_r^*(0)} = 0.
\end{align}
This is true for equilibrium systems as it is a product of a function even in time and another function odd in time. However, we assume this to be true even for non-equilibrium steady states, and leave the general case for future work.  In such a case, \eqref{eq:app-der-6} reduces to 
\begin{align}
      \brak{\dot{A}_i(\tau)  A_r^*(0)} &= - \int_0 ^{\tau} d t'' \brak{\dot{A}_i(t'')  \dot{A}_r^*(0)}. \label{eq:app-der-7}
\end{align}

Using \eqref{eq:app-der-7}, \eqref{eq:app-der-4} can be rewritten as 
\begin{align}
      M_{ij}S_{jk}\brak{A_k(0) A_r^*(0)} & = - \frac{1}{\Delta t} \Bigg[\int_0^{\Delta t}dt' \int_0 ^{t'} d \tau \brak{\dot{A}_i(\tau)  \dot{A}_r^*(0)} \Bigg] \\
      & = - \int_0^{\Delta t} d\tau  \Big(1 - \frac{\tau}{\Delta t}\Big)\brak{\dot{A}_i(\tau)  \dot{A}_r^*(0)},  \label{eq:app-der-8}
\end{align}
where the second equality is obtained by exchanging the order of integration. If the time decay $\tau_{\text{corr}}$ of the auto-correlation function of $\dot{A}_i$ and $\dot{A}_r$ is small compared to the time scale $\Delta t$, then the integral in \eqref{eq:app-der-8} can be rewritten to yield 
\begin{align}\label{eq:GK-app}
      M_{ij}S_{jk}\brak{A_k(0) A_r^*(0)} 
      & = -\int_0^{\infty} dt  \brak{\dot{A}_i(t)  \dot{A}_r^*(0)},
\end{align}
which are the Green-Kubo relations relating the transport coefficients $S_{jk}$ with the time integrals of the correlation functions of rates of the observables $A_i$. 
Since the dynamical equations or conservation laws \eqref{eq:dyn_det} is valid for every member of the ensemble, the Green-Kubo relations \eqref{eq:GK-app} can be rewritten in terms of the time correlation functions of the flux variables $B_j$ as
\begin{align}\label{eq:GK-app-flux}
M_{ij}S_{jk}\brak{A_k(0){A}^*_r(0)}&=-M_{ij}{M}^*_{rk}\int_{0}^\infty\brak{B_j(t) {B}^*_k(0)}dt.
\end{align}
It is this equation that will allow us to derive the Green-Kubo relations for the viscosity coefficients from the regression hypothesis on decay of fluctuations about the steady state of the fluid described in Appendix II.3.
An important component in being able to derive the Green-Kubo relations \eqref{eq:GK-app} and \eqref{eq:GK-app-flux} is the requirement on the separation of time scales: 
the time scale required to observe the decay of the fluctuations about the steady state $\Delta t$ is larger than the $\tau_{\text{corr}}
$ measuring the molecular relaxation processes in terms of the correlation functions of flux variables, and small compared to the time decay of the external perturbation $\tau_{\text{r}}$, i.e., 
\begin{align}
    \tau_{\text{corr}} \ll \Delta t \ll \tau_\text{r}. 
\end{align}

\subsection{IV. Correlation Functions and Isotropy}
In this section constraints on various correlation functions resulting from the assumption of isotropy. Let $\mathbf{s}$ and $\mathbf{t}$ be zero-mean random tensor fields, possibly of different rank, defined on a square box of side length $L$ with periodic boundary conditions. These have Fourier components
\begin{equation}
\mathbf{s}_\mathbf{k}=\frac{1}{L^2}\int e^{-\im \mathbf{k}\cdot\mathbf{x}}\,\mathbf{s}(\mathbf{x})\,d^2\mathbf{x}
\end{equation}
and similarly for $\mathbf{t}$. Assuming that spatial correlations decay rapidly with separation, we may expand the correlations of Fourier modes as 
\begin{align}
\brak{\mathbf{s}_\mathbf{k}\otimes\mathbf{t}_{-\mathbf{k}}}&=\frac{1}{L^4}\int e^{i\mathbf{k}\cdot(\mathbf{y}-\mathbf{x})}\brak{\mathbf{s}(\mathbf{x})\otimes\mathbf{t}(\mathbf{y})}d^2\mathbf{x}\,d^2\mathbf{y}\\
\nonumber&=\frac{1}{L^4}\int \brak{\mathbf{s}(\mathbf{x})\otimes\mathbf{t}(\mathbf{y})}d^2\mathbf{x}\,d^2\mathbf{y}+\frac{ik^r}{L^4}\int (y^r-x^r)\brak{\mathbf{s}(\mathbf{x})\otimes\mathbf{t}(\mathbf{y})}d^2\mathbf{x}\,d^2\mathbf{y}\\
\nonumber&\hspace{20pt}-\frac{k^rk^s}{2L^4}\int (y^r-x^r)(y^s-x^s)\brak{\mathbf{s}(\mathbf{x})\otimes\mathbf{t}(\mathbf{y})}d^2\mathbf{x}\,d^2\mathbf{y}+\mathcal{O}(k^3).
\end{align}
Assuming isotropy, the spatial correlators must satisfy $\brak{\mathbf{Q}_s\mathbf{s}(\mathbf{Q}\mathbf{x})\otimes\mathbf{Q}_t\mathbf{t}(\mathbf{Q}\mathbf{y})}=\brak{\mathbf{s}(\mathbf{x})\otimes\mathbf{t}(\mathbf{y})}$ for any proper rotation $\mathbf{Q}$ which acts on tensors of types $\mathbf{s}$ and $\mathbf{t}$ by $\mathbf{Q}_s$ and $\mathbf{Q}_t$. Using this relation, we can note that the Fourier-space correlator is isotropic to zeroth order in $\mathbf{k}$. To see that this is the case, consider any rotation $\mathbf{Q}$. The zeroth order term in the expansion satisfies:
\begin{align}
\int \brak{\mathbf{s}(\mathbf{x})\otimes\mathbf{t}(\mathbf{y})}d^2\mathbf{x}\,d^2\mathbf{y}&=\int \brak{\mathbf{Q}_s\mathbf{s}(\mathbf{Q}\mathbf{x})\otimes\mathbf{Q}_t\mathbf{t}(\mathbf{Q}\mathbf{y})}d^2\mathbf{x}\,d^2\mathbf{y}=\left(\mathbf{Q}_s\otimes\mathbf{Q}_t\right)\int \brak{\mathbf{s}(\mathbf{Q}\mathbf{x})\otimes\mathbf{t}(\mathbf{Q}\mathbf{y})}d^2\mathbf{x}\,d^2\mathbf{y}\\
\nonumber&=\left(\mathbf{Q}_s\otimes\mathbf{Q}_t\right)\int \brak{\mathbf{s}(\mathbf{x})\otimes\mathbf{t}(\mathbf{y})}d^2\mathbf{x}\,d^2\mathbf{y}
\end{align}
where the first equality uses the assumption of isotropy of the spatial correlators, and the third uses the rotational invariance of the integration measure. One consequence of this zeroth-order isotropy is that the correlators $\brak{\mathbf{s}_\mathbf{k}\otimes\mathbf{t}_{-\mathbf{k}}}$ vanish to zeroth order in $\mathbf{k}$ if the difference between the ranks of $\mathbf{s}$ and $\mathbf{t}$ is odd, as there are no isotropic odd rank tensors in two dimensions. If $\mathbf{s}=\mathbf{t}$, then the zeroth order term is also symmetric. In particular, this means that the autocorrelator of $\mathbf{v}_\mathbf{k}$ is proportional to the Kronecker tensor.

It will be convenient for the Green-Kubo analysis in the subsequent section to give compact notation for the lowest-order non-vanishing terms of the various correlators:
\begin{align}
\brak{v_\mathbf{k}^j v_{-\mathbf{k}}^k}&\approx \frac{1}{L^4}\int \brak{\delta v^j(\mathbf{x})\delta v^k(\mathbf{y})}d^2\mathbf{x}\,d^2\mathbf{y}:=\mu\delta_{jk}, \label{eq:mu-corr}\\
\brak{v_\mathbf{k}^jm_{-\mathbf{k}}}&\approx \frac{ik^r}{L^4}\int (y^r-x^r)\brak{\delta v^j(\mathbf{x})\delta m(\mathbf{y})}d^2\mathbf{x}\,d^2\mathbf{y}:=\im k^r\Omega_{rj}, \label{eq:vj-m-corr}\\
\brak{m_\mathbf{k}v_{-\mathbf{k}}^k}&\approx \frac{ik^r}{L^4}\int (y^r-x^r)\brak{\delta m(\mathbf{x})\delta v^k(\mathbf{y})}d^2\mathbf{x}\,d^2\mathbf{y}:=-\im k^r\Omega_{rk},\label{eq:m-vk-corr}\\  
\brak{m_\mathbf{k}m_{-\mathbf{k}}}&\approx \frac{1}{L^4}\int \brak{\delta m(\mathbf{x})\delta m(\mathbf{y})}d^2\mathbf{x}\,d^2\mathbf{y}:=\nu. \label{eq:mm-corr}
\end{align}

\subsection{V. Green-Kubo Relations for Non-equilibrium Viscous Fluids}
In what follows, we derive the Green-Kubo relations for the viscosity coefficients in terms of the stress auto-correlation functions. In Appendix II we expressed the dynamics of fluctuations about a $\mathbf{v}=\mathbf{0}$ and $m=constant$ steady state in in matrix form by
\begin{align}
\frac{d}{dt}\mathbf{A}&=\mathbf{M}\mathbf{B},\label{eq:det-GK}\\
\mathbf{B}&=\mathbf{S}\mathbf{A},\label{eq:consti-GK}
\end{align}
with the definitions
\begin{align}\label{eq:GK-defs-matrices}
\mathbf{A}&=\left[\begin{tabular}{c}
$v_\mathbf{k}^r$      \\
$m_\mathbf{k}$
\end{tabular}\right],\hspace{30pt}\mathbf{B}=\left[\begin{tabular}{c} 
$T_\mathbf{k}^{\mu\nu}$ \\
$C_\mathbf{k}^{\lambda}$ \\ 
\end{tabular}\right],\hspace{30pt}\mathbf{M}=\frac{1}{\rho_0}\left[\begin{tabular}{ccccccc}
$\im k^\nu\delta_{\mu r}$  & 0 &\\
 $-\epsilon_{\mu\nu}$  & $\im k^\lambda$ 
\end{tabular}\right],\hspace{30pt}\mathbf{S}=\left[\begin{tabular}{ccc}
  $\im\eta_{\mu\nu r\aleph}k^\aleph$ & $\gamma_{\mu\nu}$\\
0 & $\im\alpha_{\lambda \aleph}k^\aleph$
\end{tabular}\right].
\end{align}
These lead to the deterministic fluid transport equations
\begin{align}
    \frac{d}{dt}\mathbf{A}&=\mathbf{M}\mathbf{S}\mathbf{A}.
\end{align}

We now make the Onsager regression hypothesis presented in Appendix III, that is, we assume that the small \textit{stochastic} fluctuations about the steady state decay with the same transport equations in the spirit of \eqref{eq:app-reg-trans}.
If this assumption is valid, we may apply the machinery of Appendix III and the result \eqref{eq:GK-app-flux} to obtain the general Green-Kubo relations in matrix form as
\begin{align}\label{eq:GK-matrix}
\mathbf{M}\mathbf{S}\brak{\mathbf{A}\otimes{\mathbf{A}^*}}&=-\mathbf{M}\left(\int_{0}^\infty\brak{\mathbf{B}(t)\otimes {\mathbf{B}^*}(0)}dt\right){\mathbf{M}}^\dagger,
\end{align}
where the $\mathbf{M}^\dagger$ is the conjugate transpose of $\mathbf{M}$.
Plugging in the definitions of the matrices in \eqref{eq:GK-defs-matrices}, we evaluate the two sides of \eqref{eq:GK-matrix}, discarding in each matrix element all but the lowest order non-vanishing terms in $\mathbf{k}$. To this end, using the results \eqref{eq:mu-corr}-\eqref{eq:mm-corr} from Appendix IV, the left-hand side of \eqref{eq:GK-matrix} yields

\begin{align}
\text{LHS}&=\frac{1}{\rho_0}\left[\begin{tabular}{cc}
 $\im k^\nu\delta_{\mu i}$  & 0 \\
 $-\epsilon_{\mu\nu}$  & $\im k^\lambda$ 
\end{tabular}\right]\left[\begin{tabular}{cc}
  $\im\eta_{\mu\nu j\aleph}k^\aleph$ & $\gamma_{\mu\nu}$\\
0 & $\im\alpha_{\lambda \aleph}k^\aleph$
\end{tabular}\right]\left[\begin{tabular}{cc}
$\brak{v_\mathbf{k}^jv_{-\mathbf{k}}^k}$     &  $\brak{v_\mathbf{k}^jm_{-\mathbf{k}}}$    \\
 $\brak{m_\mathbf{k}v_{-\mathbf{k}}^k}$     &  $\brak{m_\mathbf{k}m_{-\mathbf{k}}}$ 
\end{tabular}\right]\\
\nonumber\\
&=\frac{1}{\rho_0}\left[\begin{tabular}{cc}
 $\im k^\nu\delta_{\mu i}$  & 0 \\
 $-\epsilon_{\mu\nu}$  & $\im k^\lambda$ 
\end{tabular}\right]\left[\begin{tabular}{cc}
  $\im\eta_{\mu\nu j\aleph}k^\aleph$ & $\gamma_{\mu\nu}$\\
0 & $\im\alpha_{\lambda \aleph}k^\aleph$
\end{tabular}\right]\left[\begin{tabular}{cc}
$\mu\delta_{jk}$     &  $\im k^r\Omega_{rj}$    \\
 $-\im k^r\Omega_{rk}$     &  $\nu$
\end{tabular}\right]\\
\nonumber\\
&=\frac{1}{\rho_0}\left[\begin{tabular}{cc}
$-k^\nu k^\aleph\eta_{i\nu j\aleph}$ &$\im k^\nu\gamma_{i\nu}$\\
$-\im k^\aleph\epsilon_{\mu\nu}\eta_{\mu\nu j\aleph}$& $-\epsilon_{\mu\nu}\gamma_{\mu\nu}$
\end{tabular}\right]\left[\begin{tabular}{cc}
$\mu\delta_{jk}$     &  $\im k^r\Omega_{rj}$    \\
 $-\im k^r\Omega_{rk}$     &  $\nu$
\end{tabular}\right]\\
\nonumber\\
&=\frac{1}{\rho_0}\left[\begin{tabular}{cc}
$-k^\nu k^\aleph\mu\eta_{i\nu k\aleph}+ k^r k^\nu\gamma_{i\nu}\Omega_{rk}$ & $\im \nu k^\nu\gamma_{i\nu}$ \\
 $ -\im \mu k^\aleph\epsilon_{\mu\nu}\eta_{\mu\nu k\aleph} +\im\epsilon_{\mu\nu}k^r\gamma_{\mu\nu} \Omega_{rk}$ &  $-\nu\epsilon_{\mu\nu}\gamma_{\mu\nu}$\label{eq:LHS}
\end{tabular}\right],
\end{align}
Next, the right-hand side of \eqref{eq:GK-matrix} to lowest order in the wave-vector $\mathbf{k}$ gives
\begin{align}
\text{RHS}&=-\frac{1}{\rho_0^2}\int_0^\infty \brak{\left[\begin{tabular}{cc}
 $\im k^\nu\delta_{\mu i}$  & 0\\
 $-\epsilon_{\mu\nu}$  & $\im k^\lambda$ 
\end{tabular}\right]\left[\begin{tabular}{cc}
   $T_\mathbf{k}^{\mu\nu}(t)T_{-\mathbf{k}}^{\rho\sigma}(0)$  &$T_\mathbf{k}^{\mu\nu}(t)C_{-\mathbf{k}}^{\omega}(0)$\\
   $C_\mathbf{k}^{\lambda}(t)T_{-\mathbf{k}}^{\rho\sigma}(0)$  &$C_\mathbf{k}^{\lambda}(t)C_{-\mathbf{k}}^{\omega}(0)$
\end{tabular}\right]\left[\begin{tabular}{cc}
$-\im k^\sigma\delta_{\rho k}$ & $-\epsilon_{\rho\sigma}$\\
0 &  $-\im k^\omega$
\end{tabular}\right]}dt\\
\nonumber\\
&\equiv -\frac{1}{\rho_0^2}\left[\begin{tabular}{cc}
$\im k^\nu\delta_{\mu i}$  & 0\\
 $-\epsilon_{\mu\nu}$  & $\im k^\lambda$ 
\end{tabular}\right]\left[\begin{tabular}{cc}
 $\tcor{TT}{\mu\nu\rho\sigma}$  &$\tcor{TC}{\mu\nu\omega}$\\
 $\tcor{CT}{\lambda\rho\sigma}$  &$\tcor{CC}{\lambda\omega}$
\end{tabular}\right]\left[\begin{tabular}{cc}
$-\im k^\sigma\delta_{\rho k}$ & $-\epsilon_{\rho\sigma}$\\
0 &  $-\im k^\omega$
\end{tabular}\right]\\
\nonumber\\
&=-\frac{1}{\rho_0^2}\left[\begin{tabular}{cc}
   $\im k^\nu\tcor{TT}{i\nu\rho\sigma}$  &  $\im k^\nu\tcor{TC}{i\nu\omega}$ \\
    $-\epsilon_{\mu\nu}\tcor{TT}{\mu\nu\rho\sigma}$  & $-\epsilon_{\mu\nu}\tcor{TC}{\mu\nu\omega}+\im k^\lambda\tcor{CC}{\lambda\omega}$
\end{tabular}\right]\left[\begin{tabular}{cc}
$-\im k^\sigma\delta_{\rho k}$ & $-\epsilon_{\rho\sigma}$\\
0 &  $-\im k^\omega$
\end{tabular}\right]\\
\nonumber\\
&=-\frac{1}{\rho_0^2}\left[\begin{tabular}{cc}
   $k^\nu k^\sigma \tcor{TT}{i\nu k\sigma}$ &  $-\im\epsilon_{\rho\sigma}k^\nu \tcor{TT}{i\nu\rho\sigma}$  \\
  $\im k^\sigma \epsilon_{\mu\nu}\tcor{TT}{\mu\nu k\sigma}$  & $\epsilon_{\mu\nu}\epsilon_{\rho\sigma}\tcor{TT}{\mu\nu\rho\sigma}$
\end{tabular}\right],\label{eq:RHS}
\end{align}
where we have introduced the definitions
\begin{align}
\tcor{TT}{\mu \nu \rho \sigma}&=\int_{0}^\infty\brak{ T^{\mu \nu}_\mathbf{k}(t) T^{\rho \sigma}_\mathbf{-k}(0)}\,dt, \\
\tcor{TC}{\mu \nu \omega}&=\int_{0}^\infty\brak{ T^{\mu \nu}_\mathbf{k}(t) C^{\omega}_\mathbf{-k}(0)}\,dt,\\
\tcor{CT}{\lambda \rho \sigma}&=\int_{0}^\infty\brak{ C^{\lambda }_\mathbf{k}(t) T^{\rho \sigma}_\mathbf{-k}(0)}\,dt,\\
\tcor{CC}{\lambda \omega}&=\int_{0}^\infty\brak{ C^{\lambda }_\mathbf{k}(t) 
C^{\omega}_\mathbf{-k}(0)}\,dt.
\end{align}

\noindent Equating the (2,2) entries of \eqref{eq:LHS} and \eqref{eq:RHS}, we find
\begin{align}\label{eq:epsilon-epsilon-gamma}
\epsilon_{ij}\gamma_{ij}&=\frac{1}{\rho_0\nu}\epsilon_{ij}\epsilon_{kl}\tcor{TT}{ijkl}.
\end{align}

\noindent Equating the (1,2) entries of \eqref{eq:LHS} and \eqref{eq:RHS} and contracting with $k^i$, we find
\begin{align}
k^ik^j\gamma_{ij}&=\frac{1}{\rho_0\nu}\epsilon_{kl}k^ik^j \tcor{TT}{ijkl}.
\end{align}
This equation holds independently for $\mathbf{k}=k\hat{e}_1$ and $\mathbf{k}=k\hat{e}_2$, yields two equations. Summing these amounts to replacing $k^i k^j$ with the tensor $\delta_{ij}$, resulting in
\begin{align}\label{eq:delta-gamma}
\delta_{ij}\gamma_{ij}&=\frac{1}{\rho_0\nu}\delta_{ij}\epsilon_{kl} \tcor{TT}{ijkl}.
\end{align}

\noindent Equating the (1,1) entries of \eqref{eq:LHS} and \eqref{eq:RHS}, we find
\begin{align}
-k^j k^l\mu\eta_{ij kl}+ k^j k^l\gamma_{il}\Omega_{jk}&=-\frac{1}{\rho_0}k^j k^l \tcor{TT}{ijkl}.
\end{align}
As before, this equation holds independently for $\mathbf{k}=k\hat{e}_1$ and $\mathbf{k}=k\hat{e}_2$, and therefore we may replace $k^jk^l$ by $\delta_{jl}$ to obtain
\begin{align}
\delta_{jl}\eta_{ij kl}- \frac{\gamma_{ij}\Omega_{jk}}{\mu}&=\frac{1}{\rho_0\mu}\delta_{jl} \tcor{TT}{ijkl}.
\end{align}
Contracting this equation with $\delta_{ik}$ and $\epsilon_{ik}$ we find the two equations
\begin{align}
\delta_{ik}\delta_{jl}\eta_{ij kl}- \frac{\delta_{ik}\gamma_{ij}\Omega_{jk}}{\mu}&=\frac{1}{\rho_0\mu}\delta_{ik}\delta_{jl} \tcor{TT}{ijkl},\label{eq:delta-delta}\\
\epsilon_{ik}\delta_{jl}\eta_{ij kl}- \frac{\epsilon_{ik}\gamma_{ij}\Omega_{jk}}{\mu}&=\frac{1}{\rho_0\mu}\epsilon_{ik}\delta_{jl} \tcor{TT}{ijkl}\label{eq:delta-epsilon}.
\end{align}

Equating the (2,1) entries of \eqref{eq:LHS} and \eqref{eq:RHS}, we find
\begin{align}
k^l\epsilon_{ij}\eta_{ijkl} -\epsilon_{ij}k^l\frac{\gamma_{ij} \Omega_{lk}}{\mu}&=\frac{1}{\rho_0\mu} k^l \epsilon_{ij}\tcor{TT}{ijkl}.\label{eq:21equate}
\end{align}
Contracting \eqref{eq:21equate} with $k^k$ and replacing $k^kk^l$ by $\delta_{kl}$, we find
\begin{align}
\epsilon_{ij}\delta_{kl}\eta_{ijkl} -\epsilon_{ij}\delta_{kl}\frac{\gamma_{ij} \Omega_{lk}}{\mu}&=\frac{1}{\rho_0\mu} \epsilon_{ij}\delta_{kl}\tcor{TT}{ijkl}.\label{eq:epsilon-delta}
\end{align}
Contracting \eqref{eq:21equate} instead with $\epsilon_{kr}k^r$ and replacing $k^rk^l$ by $\delta_{rl}$, we find
\begin{align}
\epsilon_{ij}\epsilon_{kl}\eta_{ijkl} -\epsilon_{ij}\epsilon_{kl}\frac{\gamma_{ij} \Omega_{lk}}{\mu}&=\frac{1}{\rho_0\mu}  \epsilon_{ij}\epsilon_{kl}\tcor{TT}{ijkl}.\label{eq:epsilon-epsilon}
\end{align}

Using the basis expansions \eqref{eq:gammacomponentform} and \eqref{eq:etacomponentform} of $\boldsymbol{\gamma}$ and $\boldsymbol{\eta}$ and the inner products provided in Table \ref{tab:scalars}, we may compute
\begin{alignat}{3}
\delta_{ik}\delta_{jl}\eta_{ij kl}&=2\lambda_1+4\lambda_2+2\lambda_3&&=4\beta_3,\\
\epsilon_{ik}\delta_{jl}\eta_{ij kl}&=4\lambda_4+4\lambda_5+4\lambda_6&&=4\beta_4,\\
\epsilon_{ij}\delta_{kl}\eta_{ijkl}&=8\lambda_5&&=2\beta_4+4\beta_5-2\beta_6,\\
\epsilon_{ij}\epsilon_{kl}\eta_{ijkl}&=4\lambda_3&&=-4\beta_1+2\beta_2+2\beta_3.
\end{alignat}
Now using $\gamma_{ij}=\gamma_1\delta_{ij}+\gamma_2\epsilon_{ij}$, defining $\tau=\epsilon_{ij}\Omega_{ij}$ and $\pi=\delta_{ij}\Omega_{ij}$, and using the equations \eqref{eq:epsilon-epsilon-gamma}, \eqref{eq:delta-gamma}, \eqref{eq:delta-delta}, \eqref{eq:delta-epsilon}, \eqref{eq:epsilon-delta}, \eqref{eq:epsilon-epsilon}, we finally have the six Green-Kubo equations relating the viscosity coefficients as 
\begin{align}
\gamma_1&=\frac{1}{2\rho_0\nu}\delta_{ij}\epsilon_{kl} \tcor{TT}{ijkl},\label{eq:GK1}\\
\gamma_2&=\frac{1}{2\rho_0\nu}\epsilon_{ij}\epsilon_{kl}\tcor{TT}{ijkl},\label{eq:GK2}\\
\lambda_1+2\lambda_2+\lambda_3- \frac{\gamma_1\pi}{2\mu}+\frac{\gamma_2\tau}{2\mu}&=\frac{1}{2\rho_0\mu}\delta_{ik}\delta_{jl} \tcor{TT}{ijkl},\label{eq:GK3}\\
\lambda_4+\lambda_5+\lambda_6- \frac{\gamma_1\tau}{4\mu}-\frac{\gamma_2\pi}{4\mu}&=\frac{1}{4\rho_0\mu}\epsilon_{ik}\delta_{jl} \tcor{TT}{ijkl}, \label{eq:GK4}\\
\lambda_5 -\frac{\gamma_2\pi}{4\mu}&=\frac{1}{8\rho_0\mu} \epsilon_{ij}\delta_{kl}\tcor{TT}{ijkl}, 
 \label{eq:GK5}\\
\lambda_3 +\frac{\gamma_2 \tau}{2\mu}&=\frac{1}{4\rho_0\mu}  \epsilon_{ij}\epsilon_{kl}\tcor{TT}{ijkl}\label{eq:GK6}.
\end{align}
Using the expansion of correlation function in small $\mathbf{k}$, we may write to lowest-order in $\mathbf{k}$ for the stress-stress correlation function
\begin{equation}
\tcor{TT}{ijkl}=\frac{1}{L^4}\int_{0}^\infty dt \int d^2\mathbf{x}\,d^2\mathbf{y} \brak{\delta T_{ij}(\mathbf{x},t)\delta T_{kl}(\mathbf{y},0)},
\end{equation}
where we have recalled that the $\mathbf{T}_\mathbf{k}$ are fluctuations about the steady-state mean stress tensor.

\begin{table}
	\def\arraystretch{2}
	\setlength{\tabcolsep}{.5em}
	\normalsize\begin{tabular}{|c|cccccc|cccccc|}\hline
		
		& $\mathbf{s}_1$ & $\mathbf{s}_2$& $\mathbf{s}_3$ & $\mathbf{s}_4$ & $\mathbf{s}_5$ & $\mathbf{s}_6$ & $\mathbf{e}_1$ & $\mathbf{e}_2$ & $\mathbf{e}_3$ & $\mathbf{e}_4$ & $\mathbf{e}_5$ & $\mathbf{e}_6$\\\hline
		
		$\delta_{ik}\delta_{jl}$&     $2$ & $4$ & $2$ & $0$ & $0$ & $0$ & $0$ &$0$ &$4$ &$0$ &$0$ &$0$ \\
		$\epsilon_{ik}\delta_{jl}$&    $0$ & $0$ & $0$ & $4$ & $4$ & $4$ & $0$& $0$& $0$& $4$&$0$ &$0$ \\
		$\epsilon_{ij}\delta_{kl}$&   $0$ &$0$ &$0$ &$0$ &$8$ &$0$ & $0$&$0$ &$0$ &$2$ &$4$ &$-2$ \\
		$\epsilon_{ij}\epsilon_{kl}$& $0$ &$0$ &$4$ &$0$ &$0$ &$0$ & $-4$&$2$ &$2$ &$0$ &$0$ &$0$ \\\hline
	\end{tabular}
	\caption{Inner products of the tensors appearing in the Green-Kubo relations \eqref{eq:GK3}-\eqref{eq:GK6} with the (non-normalized) elements of the two orthogonal bases for the isotropic rank four tensors (in two dimensions) introduced in Appendix I.}\label{tab:scalars}
\end{table}

\end{document}